\documentclass[showpacs,preprintnumbers,amsmath,amssymb,twocolumn]{revtex4}
\usepackage{subcaption}
\usepackage[colorlinks=true,bookmarks=true]{hyperref}
\usepackage{float}
\usepackage{adjustbox}
\usepackage{booktabs}
\usepackage{multirow}
\usepackage{hyperref}
\usepackage{epstopdf}
\usepackage{bigints}
\usepackage{threeparttable}
\usepackage{graphicx}
\usepackage{dcolumn}
\usepackage{bm}
\usepackage{color}
\usepackage{caption}
\usepackage{hyperref}
\hypersetup{
  colorlinks=true,        
  linkcolor=red,         
  citecolor=blue,         
}

\begin{document}
\title{Particle acceleration around rotating Einstein-Born-Infeld black hole and plasma effect on gravitational lensing}
\author{Gulmina Zaman Babar$^{1}$}
\email{\textcolor{cyan}{gulminazamanbabar@yahoo.com}}
\author{Farruh Atamurotov$^{2,3,4}$}
\email{\textcolor{cyan}{atamurotov@yahoo.com,}\textcolor{cyan}{f.atamurotov@inha.uz}}
\author{Shafqat Ul Islam$^{5}$}
\email{ \textcolor{cyan}{Shafphy@gmail.com}}
\author{Sushant~G.~Ghosh$^{5,6}$}
\email{\textcolor{cyan}{sghosh2@jmi.ac.in,sgghosh@gmail.com}}
\affiliation{$^{1}$School of Natural Sciences, National University of Sciences and Technology, Sector H-12, Islamabad, Pakistan\\
$^{2}$Inha University in Tashkent, Ziyolilar 9, Tashkent 100170, Uzbekistan \\
$^{3}$Akfa University, Kichik Halqa Yuli Street 17,  Tashkent 100095, Uzbekistan \\
$^{4}$Ulugh Beg Astronomical Institute, Astronomy St. 33, Tashkent 100052, Uzbekistan\\
$^{5}$Centre for Theoretical Physics,Jamia Millia Islamia, New Delhi 110025, India\\
$^{6}$ Astrophysics and Cosmology Research Unit,School of Mathematics, Statistics and Computer Science,
University of KwaZulu-Natal, Private Bag 54001, Durban 4000, South Africa}

\begin{abstract}
We consider a time-like geodesics in the background of rotating Einstein-Born-Infeld (EBI) black hole to examine
the horizon and ergosphere structure. The effective potential that governs the particle's motion in the spacetime and the innermost stable circular orbits (ISCO) is also studied. A qualitative analysis is conducted to find the redshifted ultrahigh centre-of-mass (CM) energy as a result of a two-particle collision specifically near the horizon. The recent Event Horizon Telescope (EHT) triggered a surge of interest in strong gravitational lensing by black holes, which provide a new tool comparing the black hole lensing in general relativity and alternate gravity theories.  Motivated by this, we also discussed both strong and weak-field gravitational lensing in the space-time discretely for a uniform plasma and a singular isothermal sphere. We calculated the light deflection coefficients $\bar{a}$ and $\bar{b}$ in the strong field limits, and their variance with the rotational parameter $a$ for different plasma frequency as well as in vacuum. For EBI black holes, we found that plasma's presence increases the photon sphere radius, the deflection angle, the deflection coefficients $\bar{a}$, $\bar{b}$, the angular positions and the angular separation between the relativistic images. It is also shown that with increasing spin the impact of plasma on a strong gravitational lensing becomes smaller as the spin parameter grows in the prograde orbit ($a>0$). For extreme black holes, the strong gravitational effects in the homogenous plasma are similar to those of in a vacuum. We investigate strong gravitational lensing effects by supermassive black holes Sgr A* and M87*. Considering rotating EBI black holes as the lens, we find the  angular position of images for Sgr A* and M87* and observe that the deviations of the angular position from that of the analogous Kerr black hole are not more than $2.44~\mu$as for Sgr A* and $1.83~\mu$as for M87*, which are unlikely to get resolved by the current EHT observations.
\end{abstract}

\maketitle
\section{Introduction} \label{intro}
The existence of black holes has remained a subject of interest ever since Einstein discovered their presence in the Universe
with General Relativity's theory. The black holes were believed to form when the matter in space experienced a cataclysmic collapse and eventually
reduced to a singular point, where both the density and curvature of the spacetime approach to infinity. However, this theory
could not convince the presence of black holes on observational level, because of this misconception
Roger Penrose proposed the \textit{Cosmic censorship conjecture}, which assuredly denies the existence of the naked singularities.
According to the latter theory the event horizon is an integral part of the singularity. Theoretical beliefs are whatsoever
incomplete unless backed up by the observational results, therefore the probe to visualize a black hole has
recently become possible in 2019 by the Event Horizon Telescope (EHT) collaboration \cite{aki:2019a,aki:2019b}.

Penrose unveiled that the black holes can also serve as an energy source to the matter in its surroundings \cite{Penrose:1971a}.
Quite a lot lately, Bandos, Silk, and West (BSW) explored that the Kerr black holes could provide a platform to the accelerated particles
for a collision to obtain a ultrahigh energy near the horizon \cite{Bds:2009a}. Up till now, an extensive research has been done in analogy
to the BSW mechanism for different gravities
\cite{Zasl:2010a,Grib:2011a,Zasl:2012a,Zasl:2012b,Zasl:2012c,Turs:2013a,Shay:2013a,Zasl:2013aa,Far:2013aa,Abu:2015e}. In some of the noteworthy articles \cite{Wei:2011a,Abu:2011c,Abu:2013c,Josh:2015a,Amir:2015pja,Ghosh:2014mea,Zhang:2019a} the authors studied the BSW effect in backgrounds of Kerr-Newman, Ho$\mathrm{\hat{r}}$ava-Lifshitz, five-dimensional Kerr black hole, Bardeen space-time, Kerr–de Sitter and Kerr–anti–
de Sitter black holes, correspondingly. Considering the Kerr spacetime, Harada \cite{Harada:2011a} investigated the CM energies
resulting from a collision between two particles in the ISCO. Although the controversial arguments still exist about the naked singularities, nevertheless Patil and Joshi researched the CM energies in the surroundings of naked singularities \cite{patail:2012a,patail:2012b}
and surprisingly, succeeded to achieve a large amount of energy near the singularities. Later on, Stuchl{\'\i}k et~al.
explored ultra-high energy collisions in the vicinity of Kehagias-Sfetsos gravity\cite{Stuc:2014a}.
Most recently, the CM energy generated in various spacetimes are addressed in the articles \cite{Vrba:2019a,Hejda:2019a,Ray:2020a,Abu:2020e}.

The black hole gravity allows several paranormal astrophysical activities in its surroundings, wherein gravitational lensing
is one of them. Assuming a strong-field regime Virbhadra and Ellis analysed the null-geodesics in \cite{Virbha:2000a},
which in fact made an unprecedented contribution to the exploration of black holes with a unique approach. The authors also extended their work for naked
singularities in \cite{Virbha:2002a}. Bozza et~al. employed analytical technique in \cite{Bozza:2001aa,Bozza:2002b} to study the properties
of strong-field lensing. Furthermore, some of the interesting additions which analysed this specific field efficiently are provided in \cite{Bozza:2003a,Sevb:2004a,Bozza:2005a,Bozza:2006a,Eiroa:2002b,Eiroa:2004a,Eiroa:2005a,Wei:2012b,Eiroa:2014a,Sotani:2015a,Zhao:2017a}.
So far, this event has been studied by many astrophysicists predominantly for the black holes surrounded by a plasma, however,
Bisnovatyi-Kogan and Tsupko have played a significant role in the investigations \cite{Bin:2010a,Tsp:2011a,Tsp:2014a,Tsp:2015a}. In accordance with the aforesaid analysis references
\cite{Abu:2013a,Atamurotov:2021a,Rog:2015a,Islam:2020xmy,Kumar:2020sag,Ghosh:2020spb,Far:2016a,Bis:2017a,Abu:2017aa,Abu:2017a,Car:2018a,Chak:2018a,Turi:2019a,Babar:2020a}
yield compelling details for various gravitational spacetimes.

In this paper our main interest is to examine the time-like and null geodesics in the rotating Einstein-Born-Infeld (EBI) spacetime
to retrace the BSW effect along with the strong and weak-field gravitational lensing, respectively. Before the EBI gravity came into acknowledgment,
it was Born and Infeld \cite{Born:1934a} who primarily introduced the nonlinear electromagnetic field in the classical electrodynamics
which was afterwards coupled with the general relativity by Hoffmann \cite{Hoff:1935a} to obtain a spherically symmetric solution for the
gravitational field of an electrically charged object. The EBI gravity received well acclaim
when the Born–Infeld type actions appeared consecutively \cite{Frad:1985a,Berg:1987a,Mets:1987a} in the light of low energy string theory. We can get
a productive understanding of the EBI spacetime from the research articles
\cite{Demi:1986a,Gibb:1995a,Gibb:1996b,DP:1997a,Chr:2000a,Feig:1998a,Feig:1998b,Fernan:2003a,Comel:2005a,Comel:2004a,Nieto:2004a,Wohl:2004a},
regarding its optimal solution and distinct features.

The rest of our paper is structured as follows. Sec.~(\ref{metrica}) incorporates the analysis of generic features of the black hole mainly the horizons and
the ergosphere region. In Sec.~(\ref{centerofmass}) the effective potential in tandem with the center-of-mass energy of collision between two
identical particles is thoroughly examined. Sec.~(\ref{stronglens}) sheds light on the gravitational lensing phenomenon in a strong-field regime distinctively for the vacuum and plasma backgrounds along with the corresponding observables. Sec.~(\ref{lensing}) renders an elaborate analysis of the deflection angle caused by the deviation of photons in a weak-field approximation considering two separate cases of the medium surrounding
the black hole, i.e, a uniform plasma and a singular isothermal sphere. Finally, in Sec.~(\ref{con}) we briefed about the main results.

\section{Geodesic of rotating EBI black holes}\label{metrica}
The action which leads to the field equations of EBI gravity is
obtained when the gravitational field is coupled to a nonlinear
Born–Infeld electrodynamics \cite{Born:1934a} in $(3 + 1)$ dimensions \cite{Hoff:1935a} reads

\begin{align}
I=\bigintssss d^4x\sqrt{-g}\Bigg(\frac{R}{16\pi G}+\mathcal{L}(\mathcal{F})\Bigg),  \label{actionfield}
\end{align}
\\
where $R$ is the scalar curvature and $g=\mathrm{det}|g_{\mu\nu}|$. The Lagrangian $\mathcal{L}(\mathcal{F})$ is defined as

\begin{align}
\mathcal{L}(\mathcal{F})=\frac{\beta^2}{4\pi G}\Bigg(1-\sqrt{1+\frac{2\mathcal{F}}{\beta^2}}\Bigg). \label{ell}
\end{align}
\\
$\mathcal{F}=\frac{1}{4}F_{\mu\nu}F^{\mu\nu}$, $F_{\mu\nu}$ indicates the electromagnetic field
tensor. The symbol $\beta$ is the Born–Infeld parameter, equal to the maximum value of electromagnetic field intensity.
The Einstein field equations and the electromagnetic field equations are constructed out of (\ref{actionfield}), respectively, as

\begin{align}
&R_{\mu\nu}-\frac{1}{2}g_{\mu\nu}R=kT_{\mu\nu},  \label{Einsteinfield}\\
&\nabla_\mu(F^{\mu\nu}\mathcal{L}_{,\mathcal{F}})=0.  \label{Electromagneticfield}
\end{align}
\\
The energy-momentum $T_{\mu\nu}$ reads

\begin{align}
T_{\mu\nu}=\mathcal{L}g_{\mu\nu}-F_{\mu\eta}F^{\eta}_{\nu}.
\end{align}
\\
$\mathcal{L}_{,F}$ represents the partial derivative of $\mathcal{L}$ with respect
to $\mathcal{F}$. The spacetime of a static and spherically symmetric
compact object with mass $M$ and a nonlinear electromagnetic
source in the EBI postulated theory has
been foremostly investigated by Hoffmann \cite{Hoff:1935a}. The metric for EBI space-time is expressed as \cite{Gibb:1995a,Gibb:1996b,DP:1997a,Chr:2000a}

\begin{align}\label{staticmetric}
ds^2&=-f(r)dt^2+f(r)^{-1}dr^2+r^2(d\theta^2+\sin^2\theta d\phi^2), \nonumber\\
f(r)&=\bigg(1-\frac{2GM}{r}+\frac{Q^2(r)}{r^2}\bigg),
\end{align}
\\
whereas, $Q^2(r)$ is a unique composition of the black hole's charge $Q$, the Born-Infeld
parameter $\beta$ and $r$.

\begin{align}
 Q^2(r)=\frac{2\beta^2r^4}{3}\bigg(1-\sqrt{1+\xi^2(r)}\bigg)\nonumber \\+\frac{4Q^2}{3}F\bigg(\frac{1}{4},\frac{1}{2},\frac{5}{4},-\xi^2(r)\bigg), \label{Q(r)}
\end{align}
\\
here, $F$ denotes the Gauss hypergeometric function and the parameter $\xi^2(r)$ is characterised by $\frac{Q^2}{\beta^2r^4}$.
The rotating EBI metric is obtained by applying the Newman-Janis algorithm~\cite{Newman:1965} to static spherical metric~(\ref{staticmetric}). The gravitational field of a rotating EBI black hole in the Boyer Lindquist coordinates is described
by the line element \cite{Lomb:2004a},
\\
\begin{align}\label{metric}
 ds^2&=-\bigg(\frac{\Delta-a^2 \sin^2\theta}{\rho^2}\bigg)dt^2+\frac{\rho^2}{\Delta}dr^2+\rho^2d\theta^2
\nonumber \\& +\sin^2\theta\bigg(\rho^2
+a^2\sin^2\theta\bigg(2-\frac{\Delta-a^2 \sin^2\theta}{\rho^2}
\bigg)\bigg)d\phi^2
\nonumber \\& -2a\sin^2\theta\bigg(1-\frac{\Delta-a^2 \sin^2\theta}{\rho^2}\bigg)dt
d\phi,
\end{align}
\begin{figure*}[t]
 \begin{center}
   \includegraphics[width=.45\textwidth]{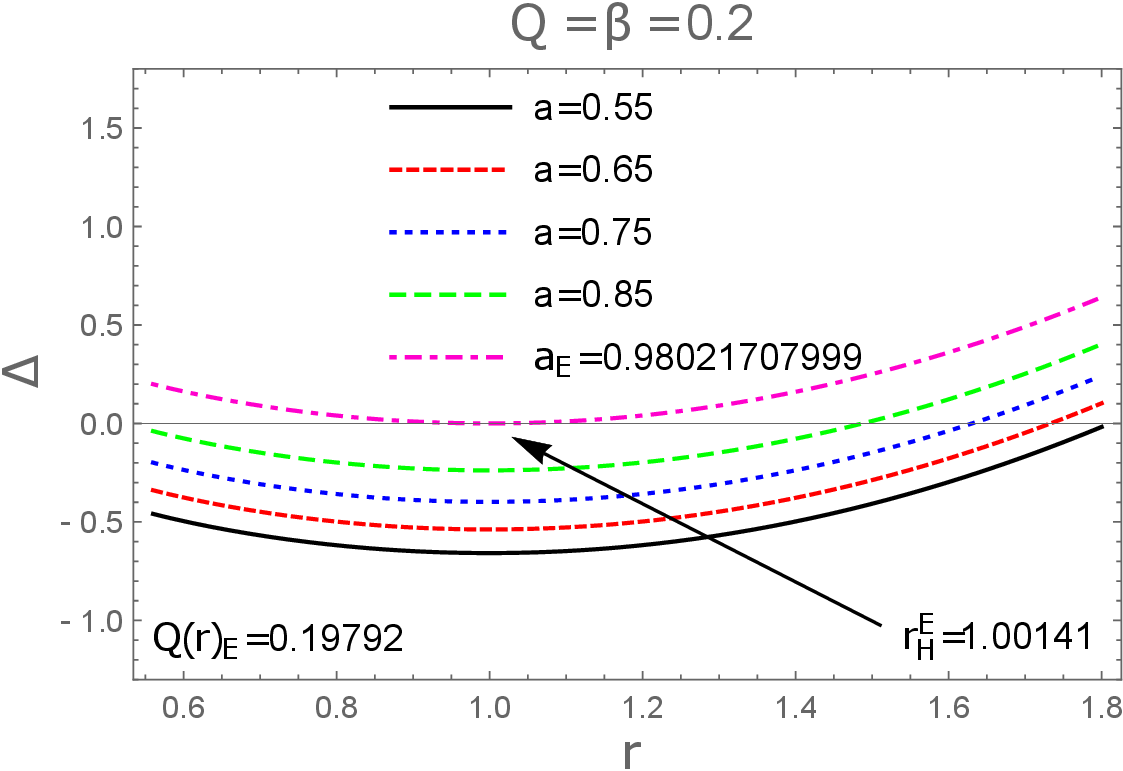}
   \includegraphics[width=.45\textwidth]{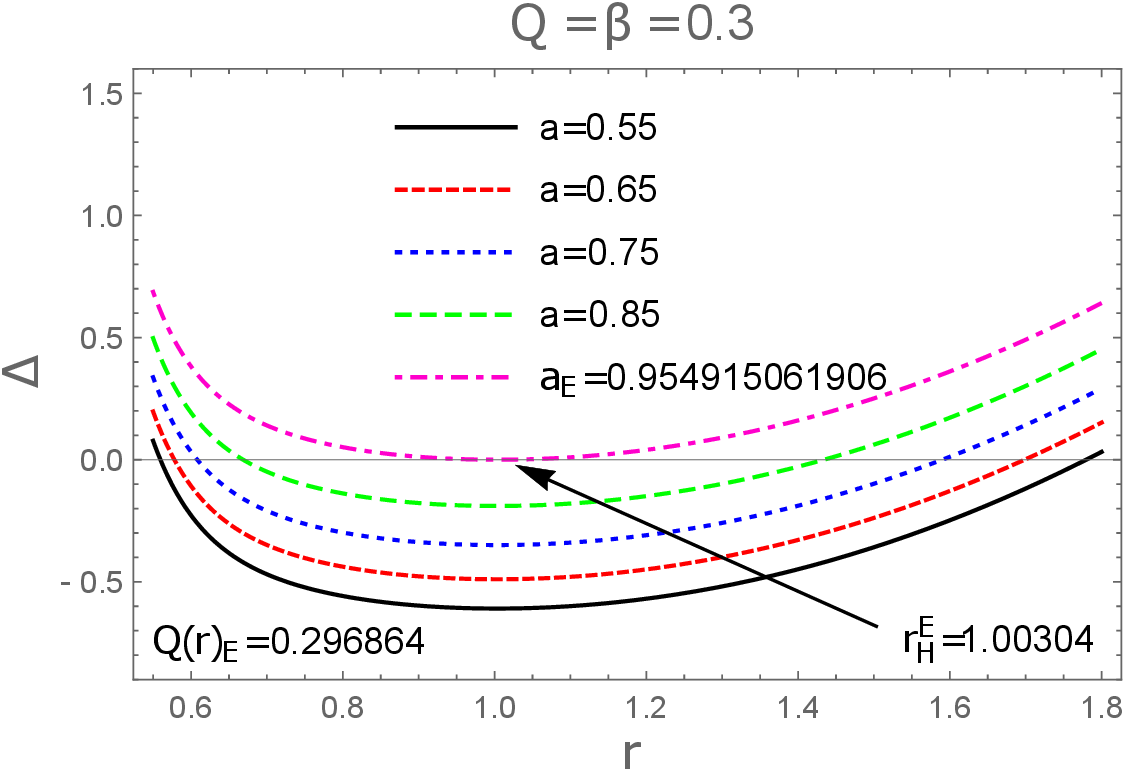}

   \includegraphics[width=.45\textwidth]{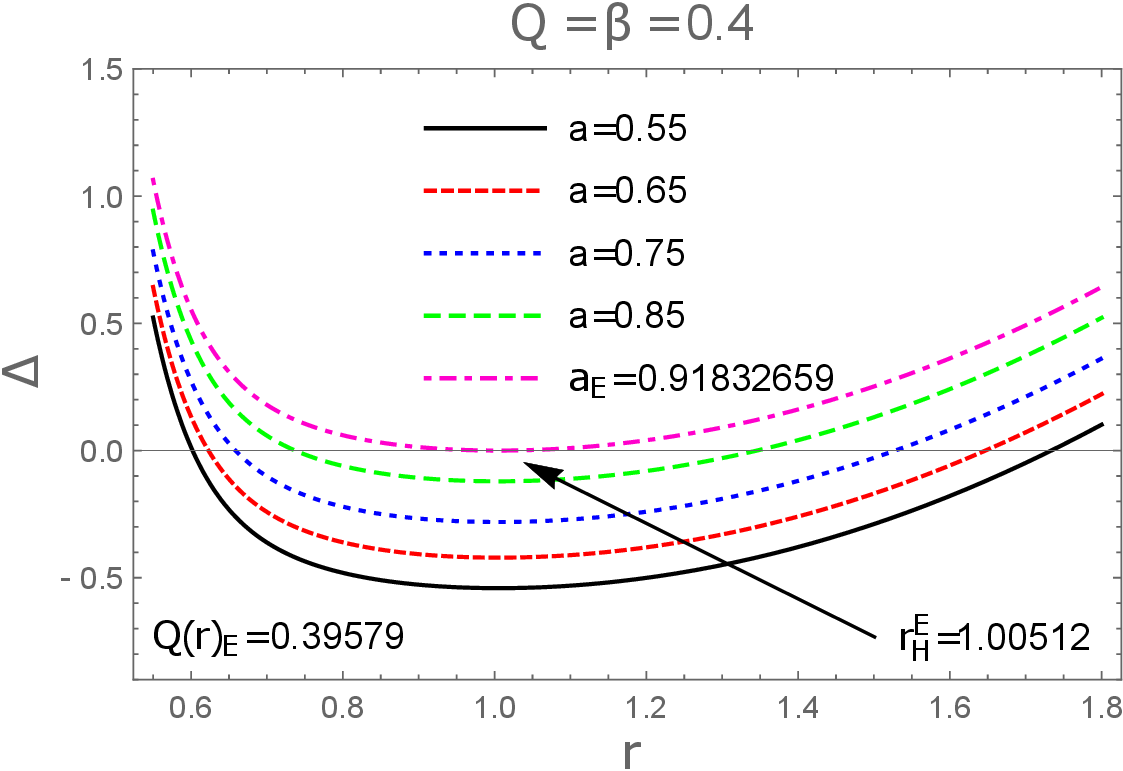}
   \includegraphics[width=.45\textwidth]{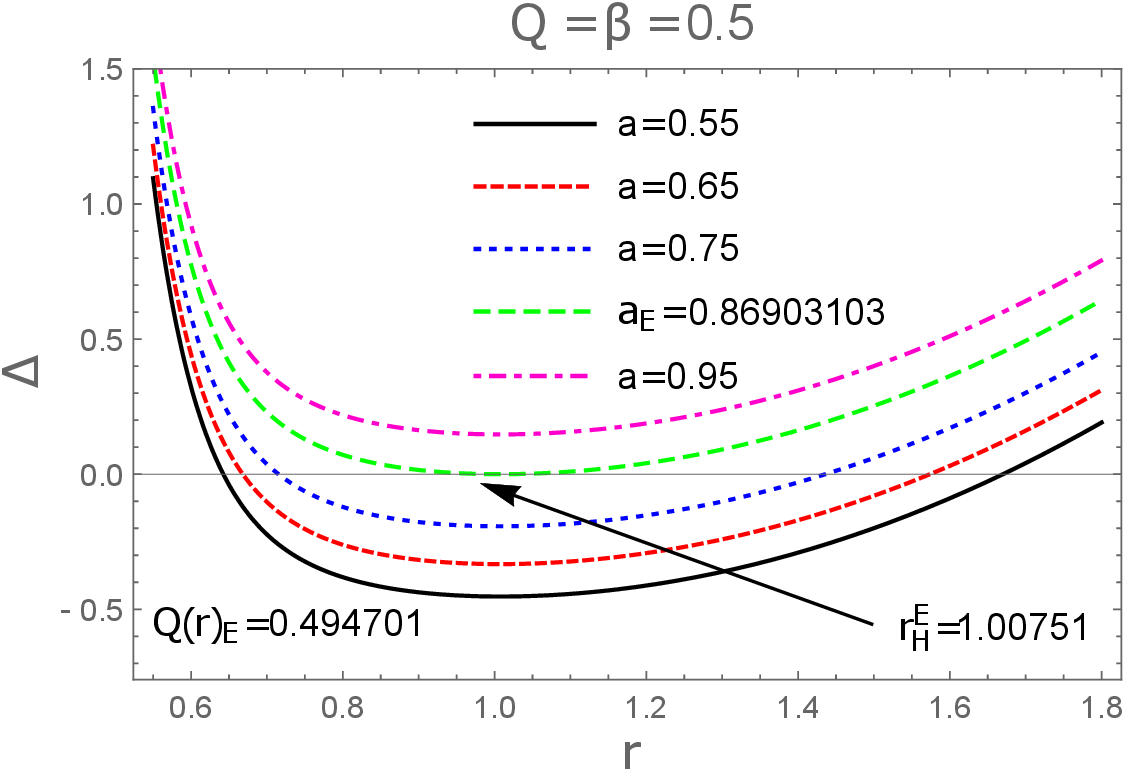}

   \includegraphics[width=.45\textwidth]{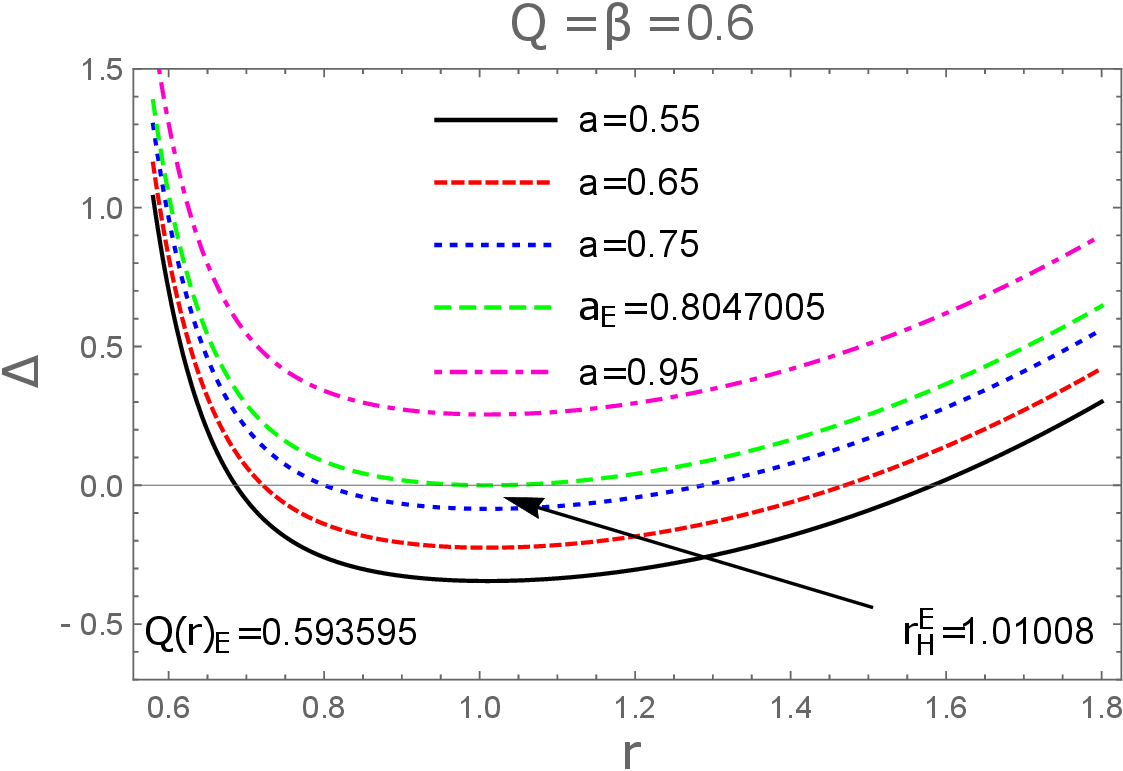}
   \includegraphics[width=.45\textwidth]{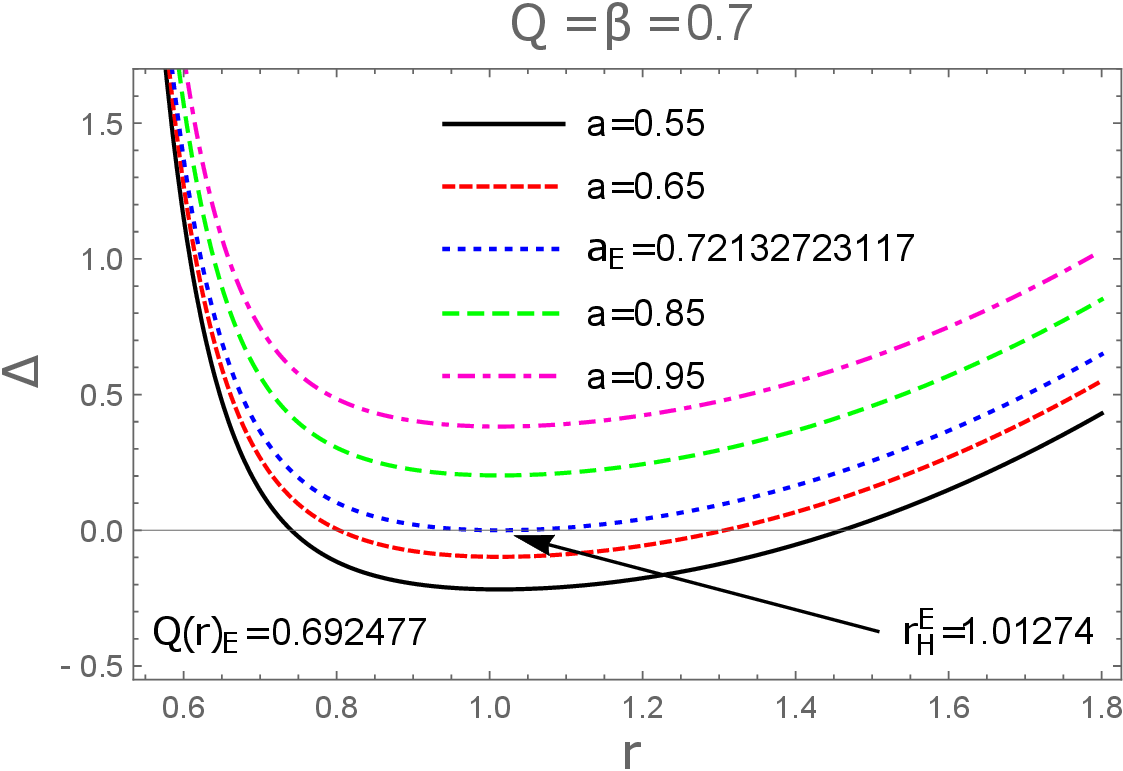}
   \end{center}
 \caption{Plot showing the behavior of $\Delta$ with respect to $r$ for different values of $Q=\beta$. The case $a = a_{E}$ corresponds to an extremal black hole.}\label{Delta}
\end{figure*}

\begin{figure*}[t]
 \begin{center}
   \includegraphics[scale=0.3]{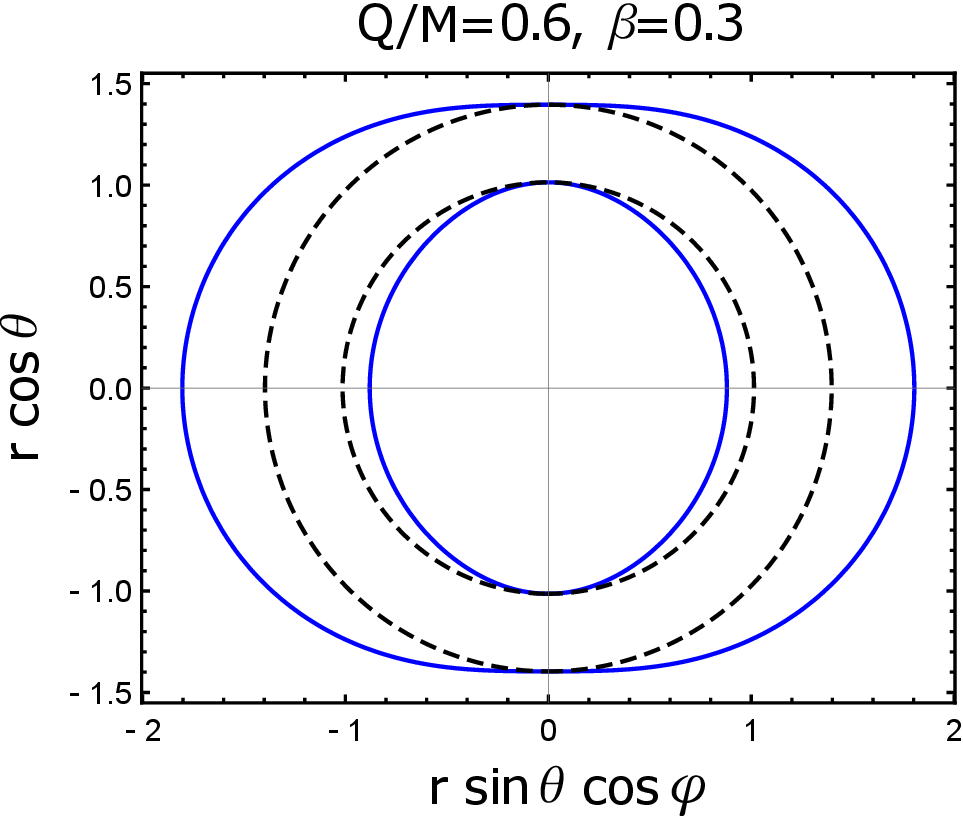}
   \includegraphics[scale=0.3]{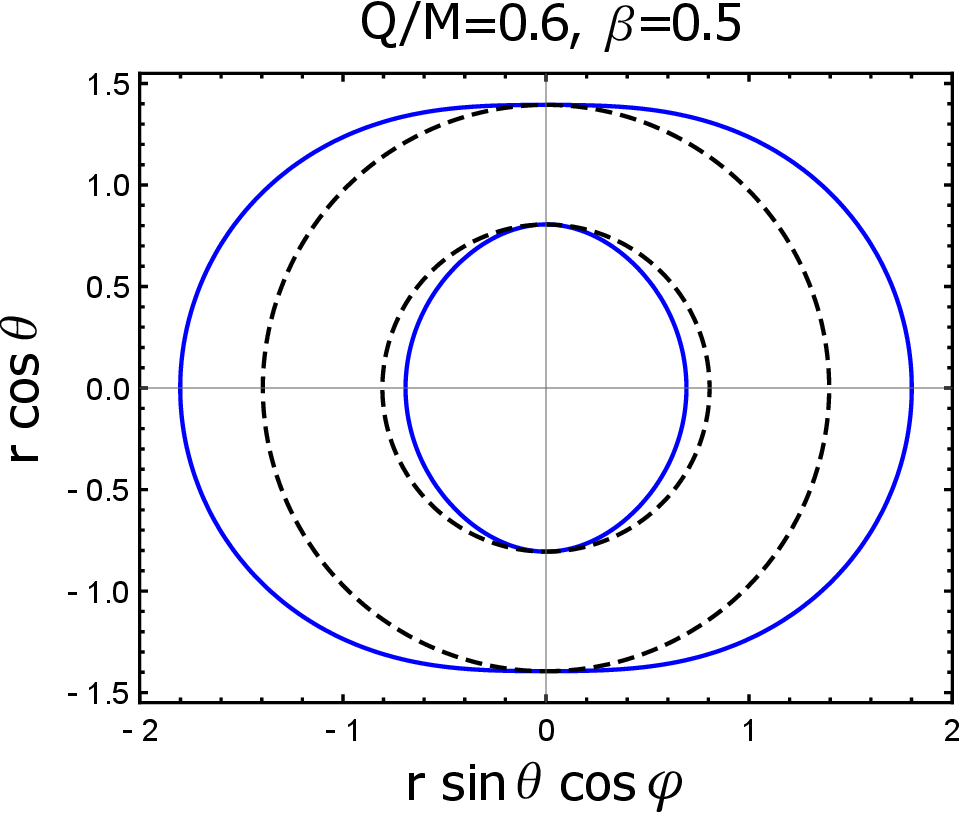}
   \includegraphics[scale=0.3]{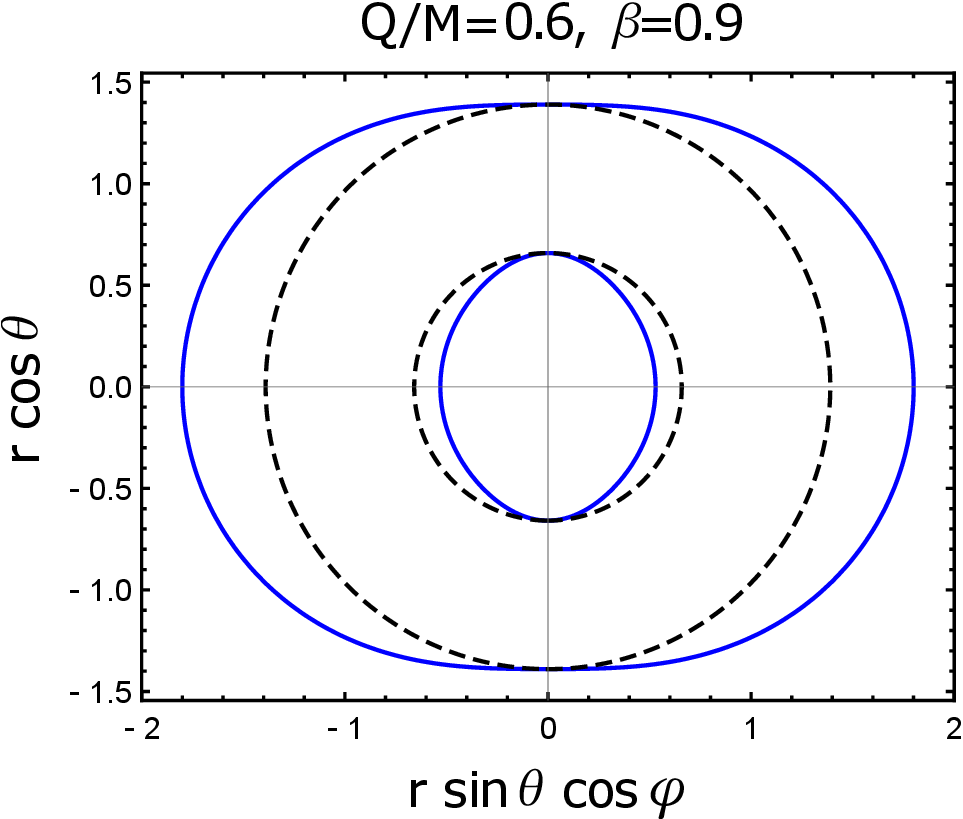}

   \includegraphics[scale=0.3]{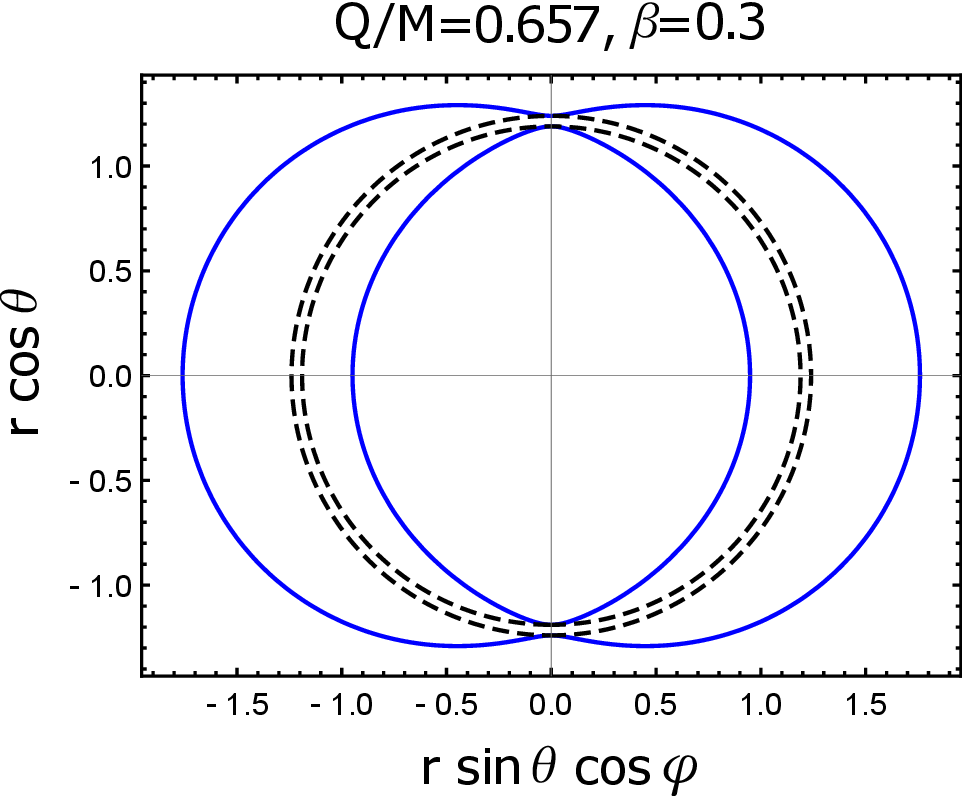}
   \includegraphics[scale=0.3]{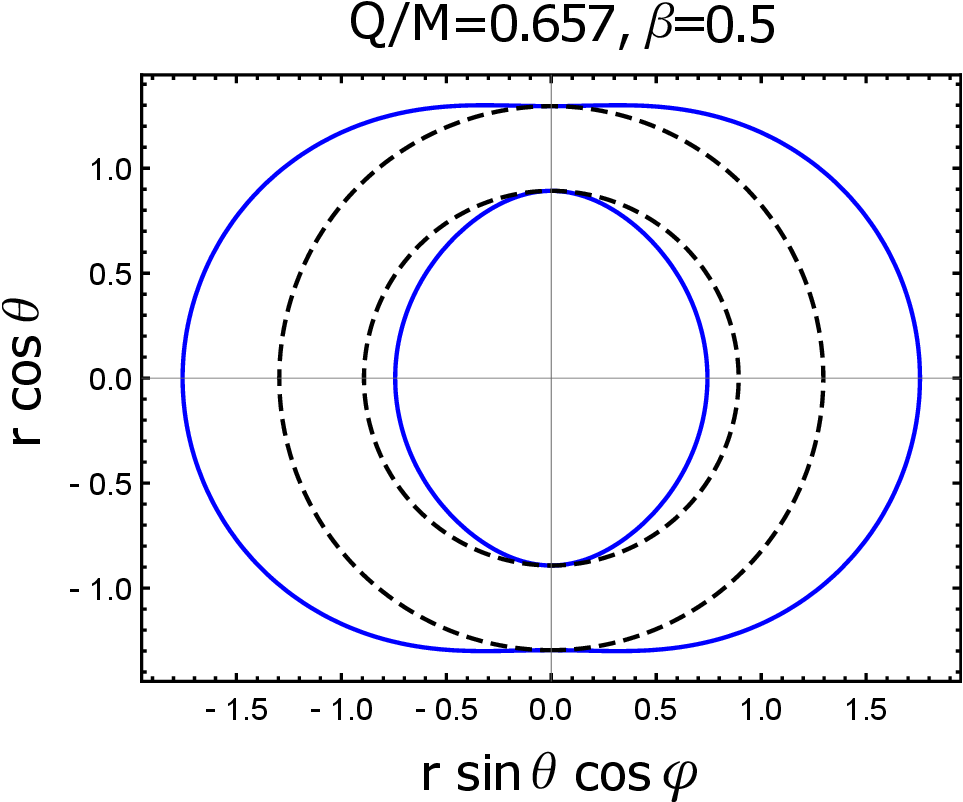}
   \includegraphics[scale=0.3]{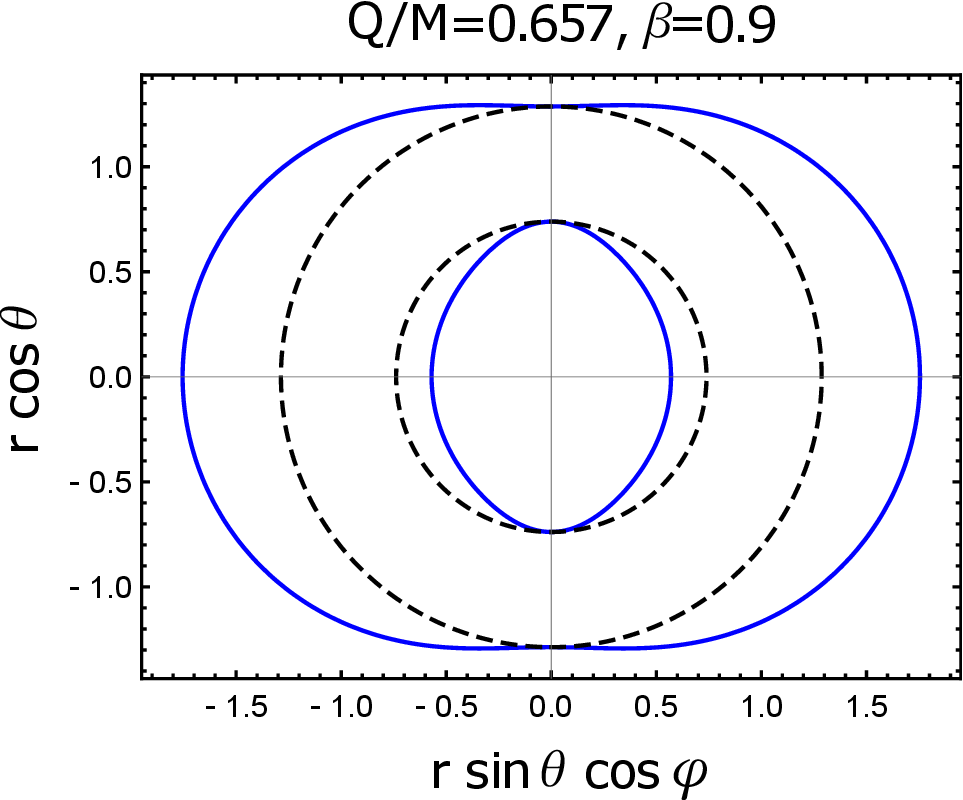}

   \includegraphics[scale=0.3]{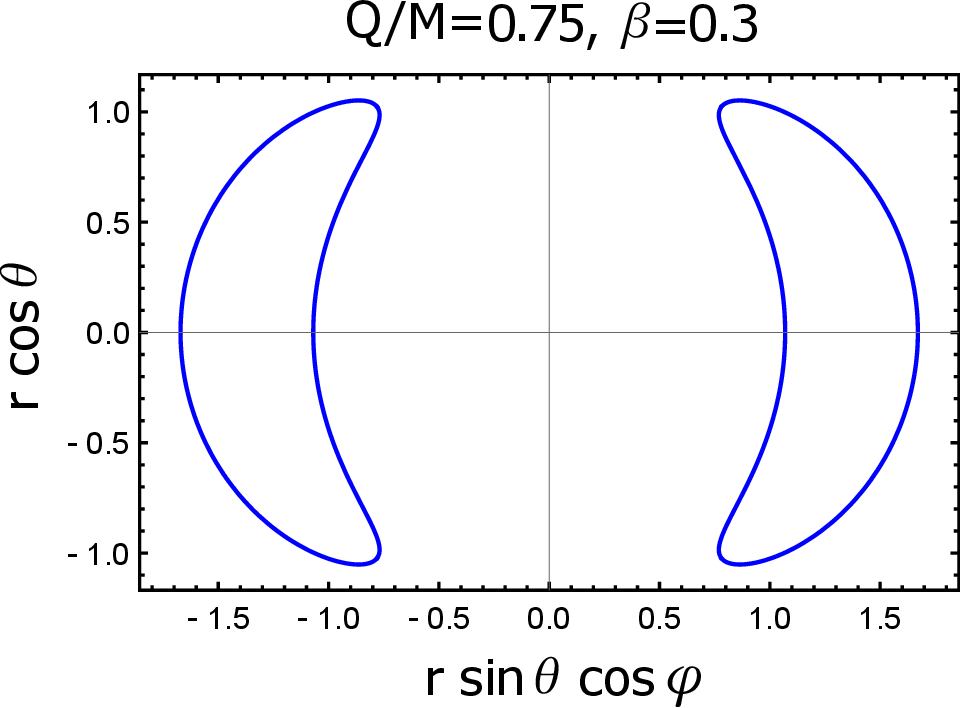}
   \includegraphics[scale=0.3]{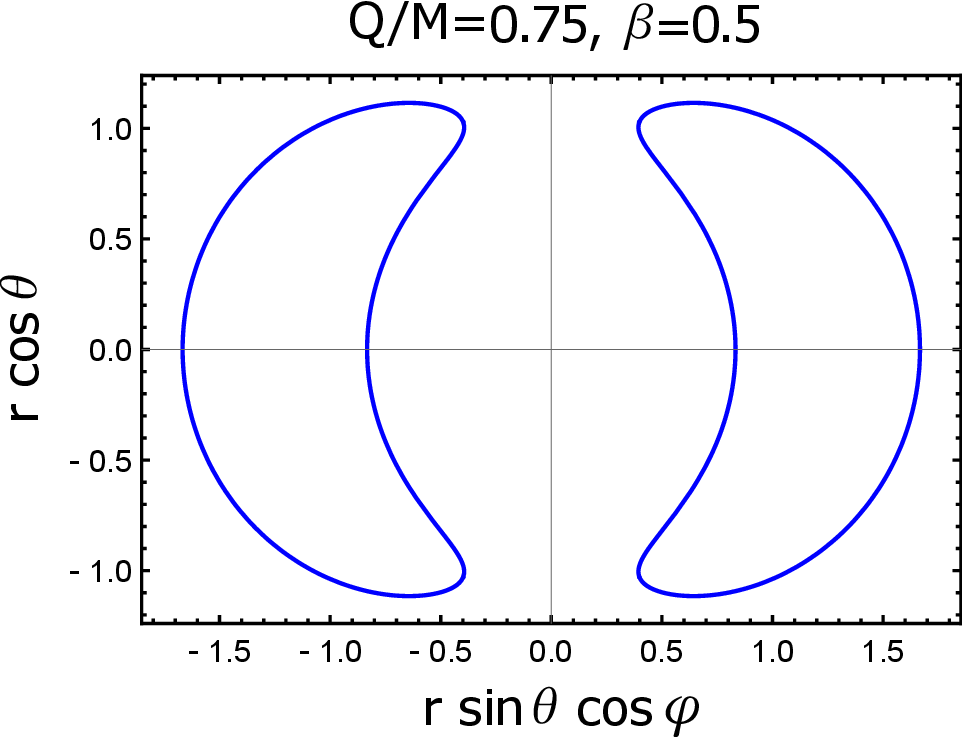}
   \includegraphics[scale=0.3]{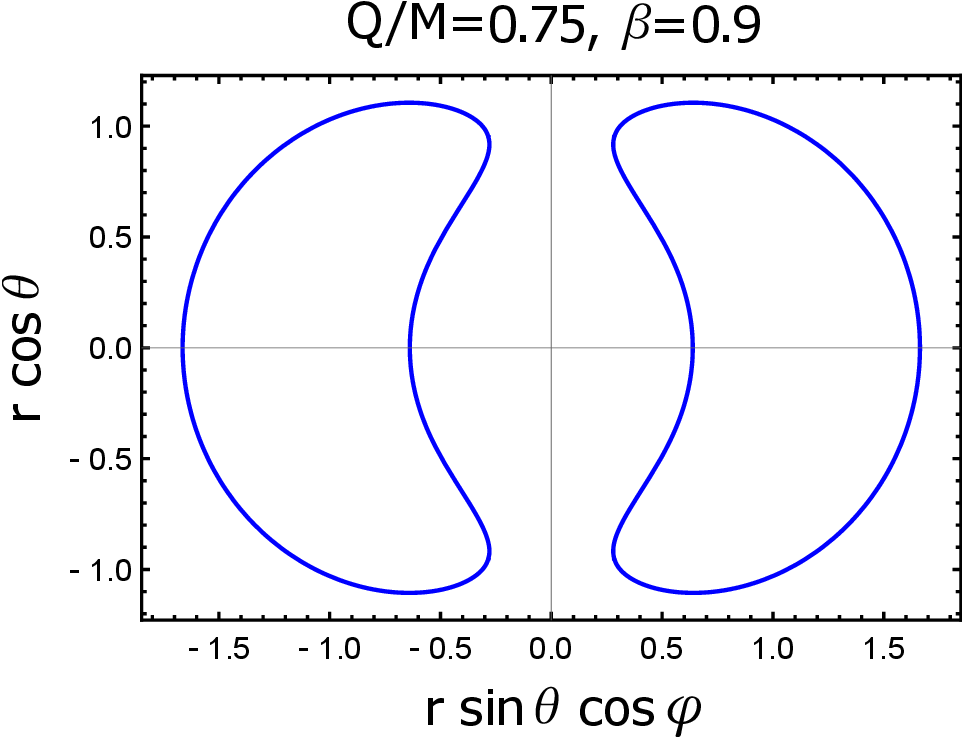}
\end{center}
\caption{Plot showing the variation of the shape of ergosphere for a rotating
charged black hole in $xz$-plane with a fixed spin parameter $a=0.7$ and for various charge $Q$ and Born-Infeld parameters $\beta$. The blue and the black lines correspond, respectively, to the static limit surfaces and horizons.}\label{Ergoevent}
\end{figure*}

\begin{figure*}[t]
 \begin{center}
 \includegraphics[scale=0.3]{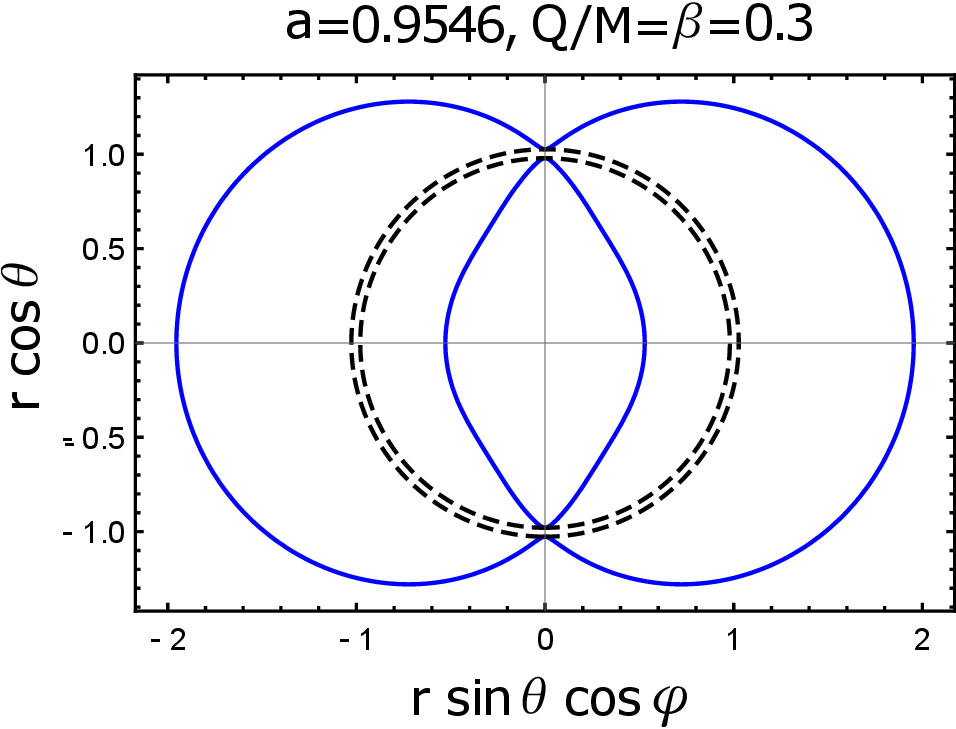}
 \includegraphics[scale=0.3]{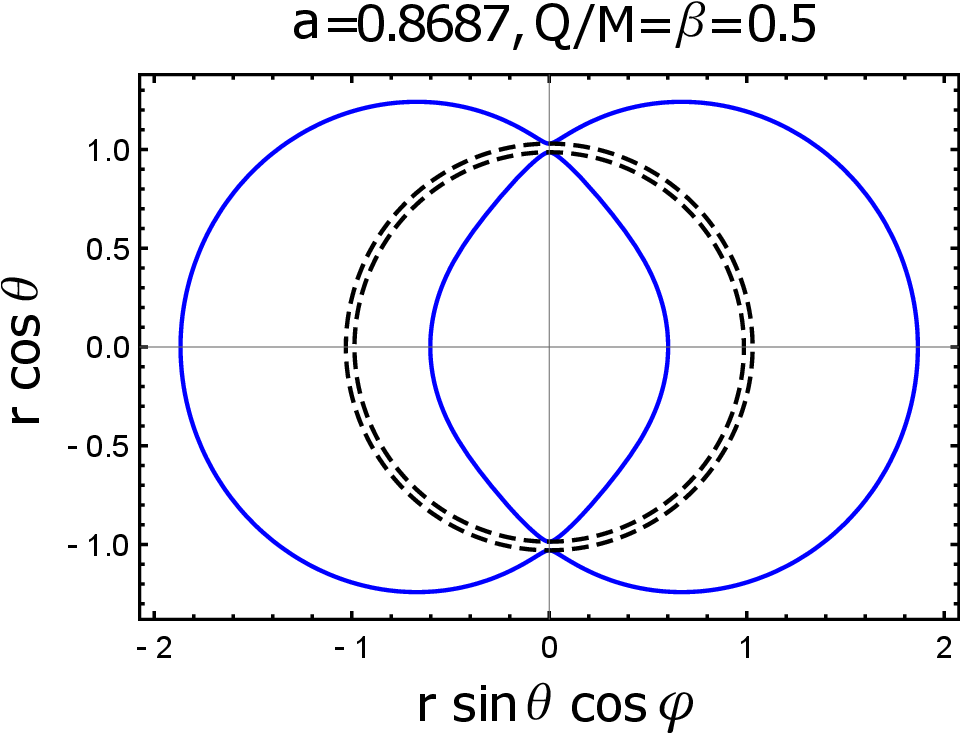}
 \includegraphics[scale=0.3]{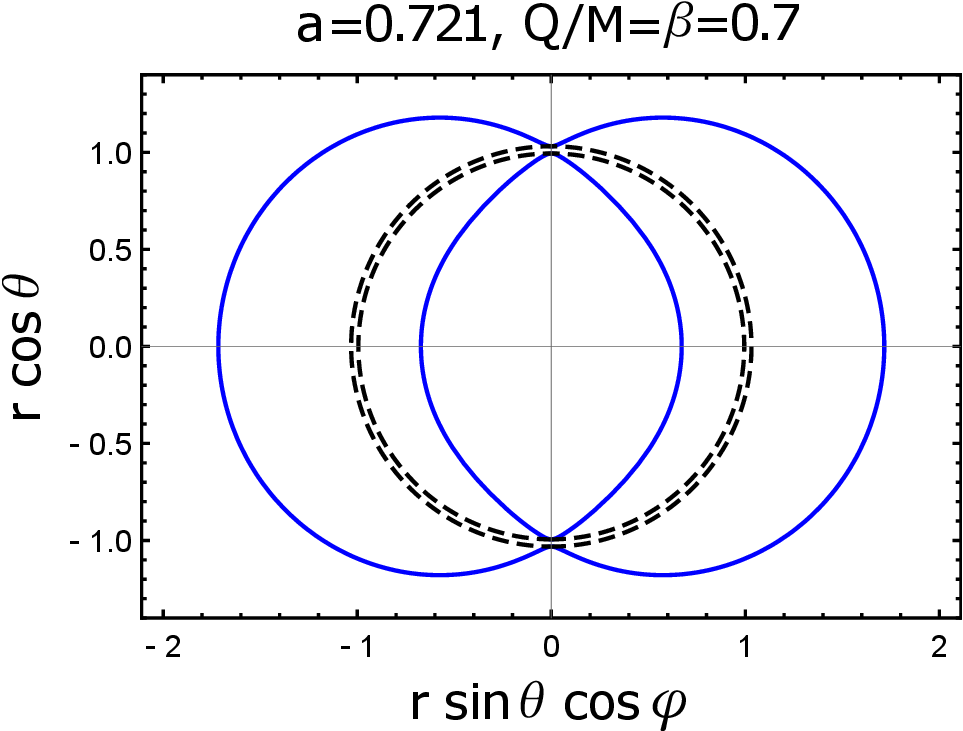}
   \end{center}
\caption{Plot showing the variation of the shape of ergosphere
for $a\approx a_E$ (extremal black hole) in $xz$-plane with parameters $Q/M=\beta$, for a rotating
 EBI black hole.}\label{ergoeventa}
\end{figure*}

\begin{table*}[t]
\caption{The values of the horizons of a rotating EBI black hole with $a$=0.5. The Kerr-Newman case is shown by $\beta\rightarrow \infty.$}\label{Tabel1}
 \resizebox{0.8\textwidth}{!}{
\begin{tabular}{@{}rrrrrrcrrrrrrrrrcrrrrr@{}}
\hline \hline
Q & \multicolumn{3}{c}{$\beta=0.4$}& \phantom{abc} & \multicolumn{3}{c}{$\beta=0.6$} & \phantom{abc}& \multicolumn{3}{c}{$\beta=0.8$} &
\phantom{abc} & \multicolumn{3}{c}{$\beta=1$} &\phantom{abc} & \multicolumn{3}{c}{$\beta \rightarrow \infty$} \\
\cmidrule[0.3pt]{2-4} \cmidrule[0.3pt]{6-8} \cmidrule[0.3pt]{10-12}  \cmidrule[0.3pt]{14-16} \cmidrule[0.2pt]{18-20}
 & $r_{H}^{-}$ && $r_{H}^{+}$  &&  $r_{H}^{-}$ && $r_{H}^{+}$
  && $r_{H}^{-}$ && $r_{H}^{+}$  && $r_{H}^{-}$ && $r_{H}^{+}$ && $r_{H}^{-}$ && $r_{H}^{+}$   \\
 \hline
0.4 & 0.5945  && 1.7686   &&0.4977  && 1.7684  && 0.4403 && 1.7683 &&0.4012 &&1.7682 && 0.2319 && 1.7681   \\
0.5 & 0.7000  && 1.7086   &&0.5851  && 1.7078  &&0.5174  && 1.7075 &&0.4717 &&1.7074 &&0.2929  && 1.7071 \\
0.6 & 0.8085  && 1.6285   &&0.6746  &&1.6265   && 0.5969 &&1.6256  &&0.5451 &&1.6252 &&0.3755  &&1.6245 \\
0.7 & 0.9289  && 1.5197   &&0.7728  &&1.5154   &&0.6849  && 1.5132 &&0.6275 &&1.5121 &&0.4901  &&1.5099   \\
0.8 & 1.1001  && 1.3452   &&0.8984  && 1.3503  &&0.7985  &&1.3443  &&0.7378 &&1.3404 &&0.6683  && 1.3316 \\
\hline\hline
\end{tabular}}
\end{table*}

\begin{table*}[t]
\caption{\small{The values of the outer static limit surface and event horizon of a rotating
EBI black hole with $a$=0.5 and $\theta=\frac{\pi}{2}$ ($\delta_{e}^{g}$= $r_{sls}^{+}$ - $r_{H}^{+}$)}. The Kerr-Newman case is shown by $\beta \rightarrow \infty$.}\label{Tabel2}
  \resizebox{1\textwidth}{!}{
\begin{tabular}{@{}rrrrrrcrrrrrrrrrcrrrrr@{}}
\hline\hline
Q & \multicolumn{3}{c}{$\beta=0.4$}& \phantom{abc} & \multicolumn{3}{c}{$\beta=0.6$} & \phantom{abc}& \multicolumn{3}{c}{$\beta=0.8$} &
\phantom{abc} & \multicolumn{3}{c}{$\beta=1$} &\phantom{abc} & \multicolumn{3}{c}{$\beta \rightarrow \infty$} \\
\cmidrule[0.3pt]{2-4} \cmidrule[0.3pt]{6-8} \cmidrule[0.3pt]{10-12}  \cmidrule[0.3pt]{14-16} \cmidrule[0.3pt]{18-20}
 & $r_{H}^{+}$ & $r_{sls}^{+}$ & $\delta_{e}^{g}$  && $r_{H}^{+}$ & $r_{sls}^{+}$ & $\delta_{e}^{g}$
  && $r_{H}^{+}$ & $r_{sls}^{+}$ & $\delta_{e}^{g}$ && $r_{H}^{+}$ & $r_{sls}^{+}$ & $\delta_{e}^{g}$ && $r_{H}^{+}$ & $r_{sls}^{+}$ & $\delta_{e}^{g}$\\
\hline

0.4 &  1.7686 & 1.9168  &0.1482  && 1.7684 & 1.9167 &0.1483  && 1.7683 &1.9166  &0.1483  &&1.7682 &1.9166 &0.1484 &&1.7681 &1.9165  &0.1484  \\
0.5 &  1.7086 & 1.8669  &0.1583  && 1.7078 &1.8664  &0.1586  && 1.7075 &1.8663  &0.1588  &&1.7074 &1.8662 &0.1588 && 1.7071&1.8660  &0.1589 \\
0.6 &  1.6285 &1.8022   &0.1737  && 1.6265 &1.8010  &0.1745  && 1.6256 & 1.8006 &0.1750  &&1.6252 &1.8004 &0.1752 && 1.6245&1.8000  &0.1755 \\
0.7 &  1.5197 &1.7192   &0.1995 && 1.5154 & 1.7167 &0.2013  && 1.5132 &1.7156   &0.2024  &&1.5121 &1.7151 &0.2030 && 1.5099 &1.7141 &0.2042  \\
0.8 &  1.3452 &1.6111   &0.2659  && 1.3503 &1.6063  &0.2560  && 1.3443 &1.6038  &0.2595  &&1.3404 &1.6025 &0.2621&& 1.3316 &1.6000  &0.2684\\
\hline\hline
\end{tabular}}
\end{table*}

with $\Delta=r^2-2Mr+a^2+Q^2(r)$ and $\rho^2=r^2+a^2\cos^2\theta$. The term $a$ refers to the spin of the black hole. The behaviour of metric~(\ref{metric}) is typical of a rotating charged source~ \cite{Lomb:2004a} and the limit $a \to 0 $ is the corresponding metric~(\ref{staticmetric}).
The imposition of certain specific restrictions on $Q(r)$ and $\beta$ leads to regenerate some well known gravities.
The Kerr-Newman gravity is obtained when $\beta\rightarrow\infty$ and $Q(r)=Q$ ($Q\neq0$)
and the solution of the Kerr black hole's is recovered when $\beta\rightarrow0$ \cite{Kerr:1963a}. Furthermore, for a static EBI metric
one can conveniently obtain the Schwarzschild and the Reissner–Nordstr$\mathrm{\ddot{o}}$m cases by setting $Q=0$ and $\beta\rightarrow\infty$, respectively. The black hole's  core where gravity tends to infinity (\textit{Curvature Singularity}) exists at $\rho=0$ and $M=Q\neq0$.
Since $0<Q<1$ and $\beta$ can take any positive real value, therefore we shall keep our discussion compact by considering
the particular case $Q=\beta$ in most of the forthcoming context to get a better understanding of the EBI spacetime.

\subsection{Horizons and ergosphere}
The EBI gravity is investigated to possess the horizon structure and the ergosphere region,
likewise the other rotating black holes. We aim to examine the properties of the above indicated features 
depending on the charge $Q$ and the Born-Infeld parameter $\beta$ \cite{Josh:2015a,Far:2016c}. The radii of the
\textit{Cauchy horizon} $r_{H}^{-}$ and the \textit{event horizon} $r_{H}^{+}$ are attained by $\Delta=0$, the coordinate singularity.
The black hole turns out to be an \textit{extremal} black hole when the two horizons coincide for a specific critical spin parameter
$a=a_E$, whereas  $a<a_E$ refers to a \textit{non-extremal} black hole with two distinct horizons. Fig.~(\ref{Delta}) shows the behaviour
of horizons by varying the spin parameter $a$ for fixed $Q=\beta$. A naked singularity is also seen to exist when $a>a_E$, because no
horizon appears in that particular case. Moreover, the black hole admits two static limit surfaces $r_{sls}^{-}$ and $r_{sls}^{+}$, which
are the positive real roots of equation $g_{tt}=0$. The region, i.e, $r_{sls}^{+}<r<r_{H}^{+}$  connotes the \textit{ergosphere} region and
it's boundary $r_{sls}^{+}$ is called the \textit{quantum ergosphere}. To evaluate the horizons and
the static limit surfaces of the EBI black hole, we shall carry out the numerical computation of $\Delta=0$ and $g_{tt}=0$,
by taking the series expansion of (\ref{Q(r)}) upto order $O(\frac{1}{\beta^5})$, given as below
\\
\begin{align}
Q^2(r)&\approx\frac{2\beta^2r^4}{3}\bigg(1-\sqrt{1+\xi^2(r)}\bigg)
\nonumber \\&+\frac{4 Q^2}{3}\bigg(1 -\frac{1}{10}\xi^2(r)+\frac{1}{24}\xi^4(r)\bigg). \label{Q(r)a}
\end{align}
\\
Fig.~(\ref{Ergoevent}) shows the evolution of the horizons and ergosphere region for a rotating EBI gravity. With the gradual charge supplement the typical ergosphere transforms considerably to a prolate shape unless the charge reaches a critical point $Q_E$,
where the two surfaces merge into one and eventually vanishes for $Q>Q_E$.
In Fig.~(\ref{ergoeventa}) the ergosphere is illustrated for the spin parameter $a$ which
is equivalent to the critical spin parameter, $a \approx a_E$. It is observed that the area becomes
larger for the greater values of $a$, therefore an accelerating charged black hole has a more prolate ergosphere \cite{Josh:2020a}.
\\
Tables (\ref{Tabel1})and (\ref{Tabel2}) hold specific data regarding the radii of the horizons and the static limit
surfaces $r^+_{sls}$ for various charge $Q$ and Born-Infeld $\beta$ parameters. An overview of the data for a fixed
Born-Infeld parameter $\beta$ with a regular charge increment manifest a uniform decrease for each of the
$r^+_{H}$ and $r^+_{sls}$, however, the Cauchy horizon rather has an absolute-counter effect. On the other hand,
scaling up the $\beta$ parameter for a specified $Q$ reduces the radii
of both the horizons along with the $r^+_{sls}$. Overall, least size radii are detected in case of the Kerr-Newman gravity.
The influence of charge on the ergosphere region is significant as compared to the $\beta$ parameter.

\section{Particle acceleration near EBI black holes}\label{centerofmass}
In this section we made an inclusive analysis to probe the acceleration of particles in the EBI gravity background. We shall precisely study the CM energy produced due to a two-particle collision near the horizon considering an extremal and a non-extremal charged
black hole. We put forth the scenario where two nonrelativistic particles initially located at infinity
fall freely towards the black hole and ultimately encounter a massive collision near the horizon. Here,
we made a unique choice for the collision point because particles falling in from infinity appear with
an infinite blueshift at the horizon and hence are considered to produce an arbitrarily large amount of energy \cite{Bds:2009a}.

We consider motion of a time-like particle with a rest mass $m_0$ in the equatorial plane $\theta=\pi/2$ where the polar velocity,
$\dot{\theta}$ becomes zero. The generalized momenta of the particle in the spacetime of a rotating charged black hole is expressed in the form,

\begin{figure*}[t]
 \begin{center}
   \includegraphics[scale=0.4]{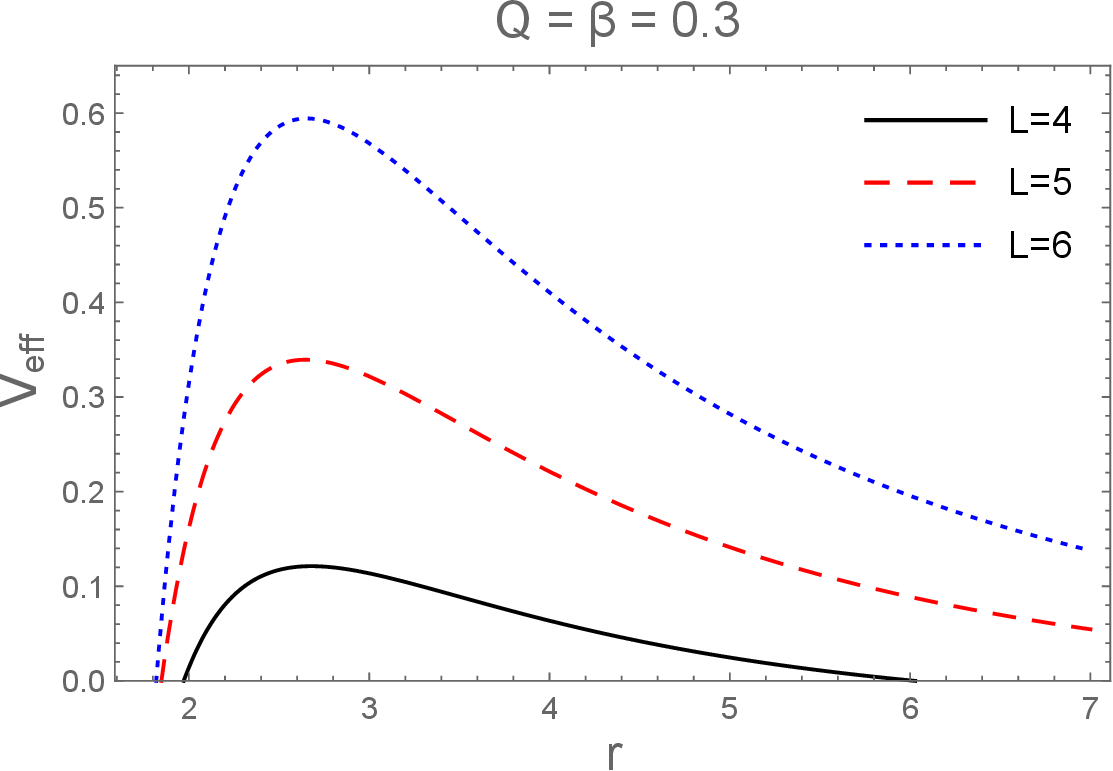}
   \includegraphics[scale=0.4]{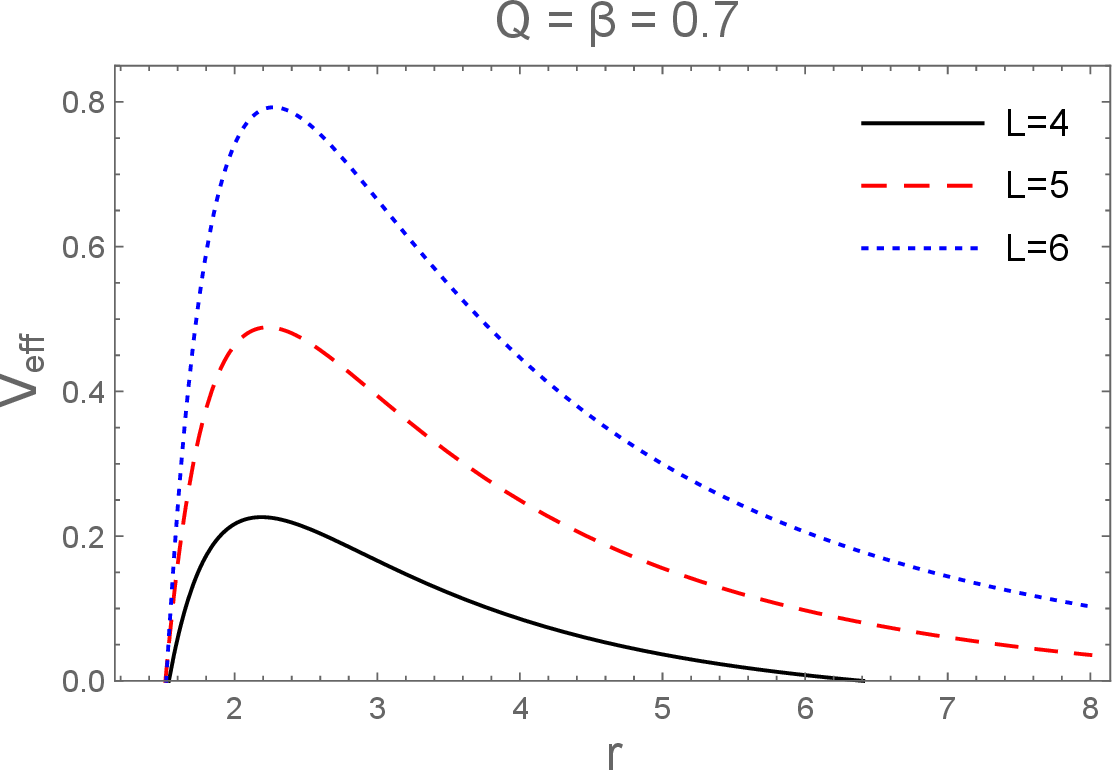}
    \includegraphics[scale=0.4]{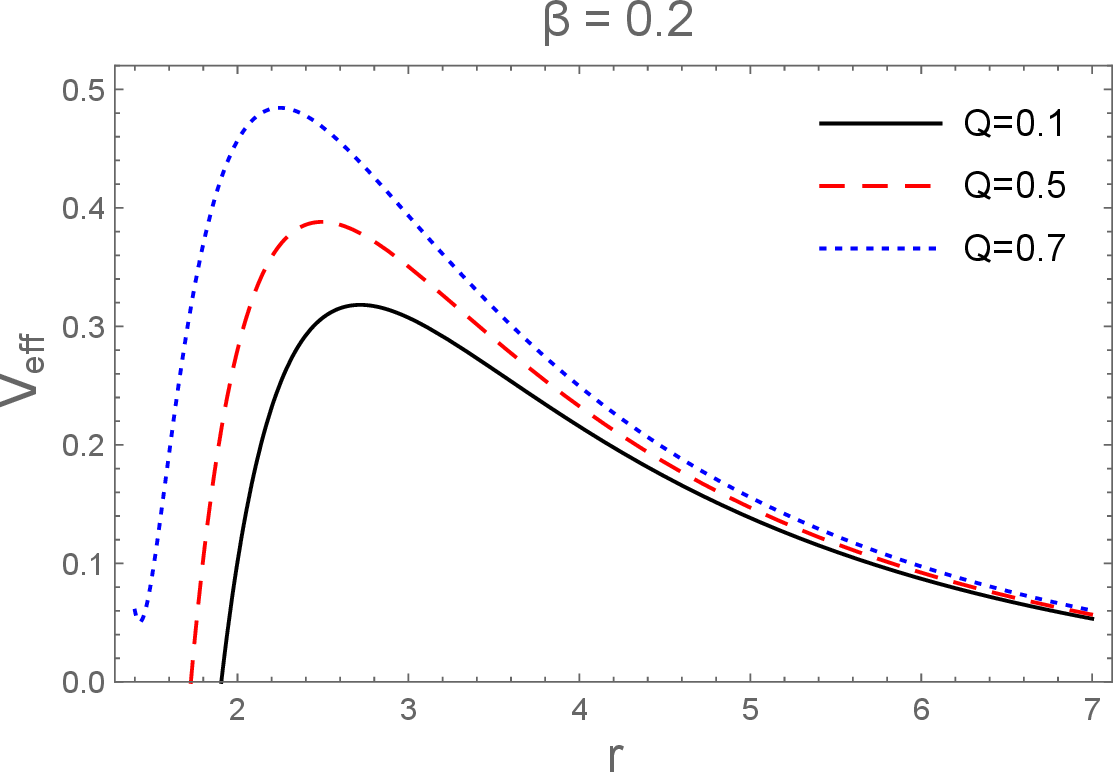}
   \includegraphics[scale=0.4]{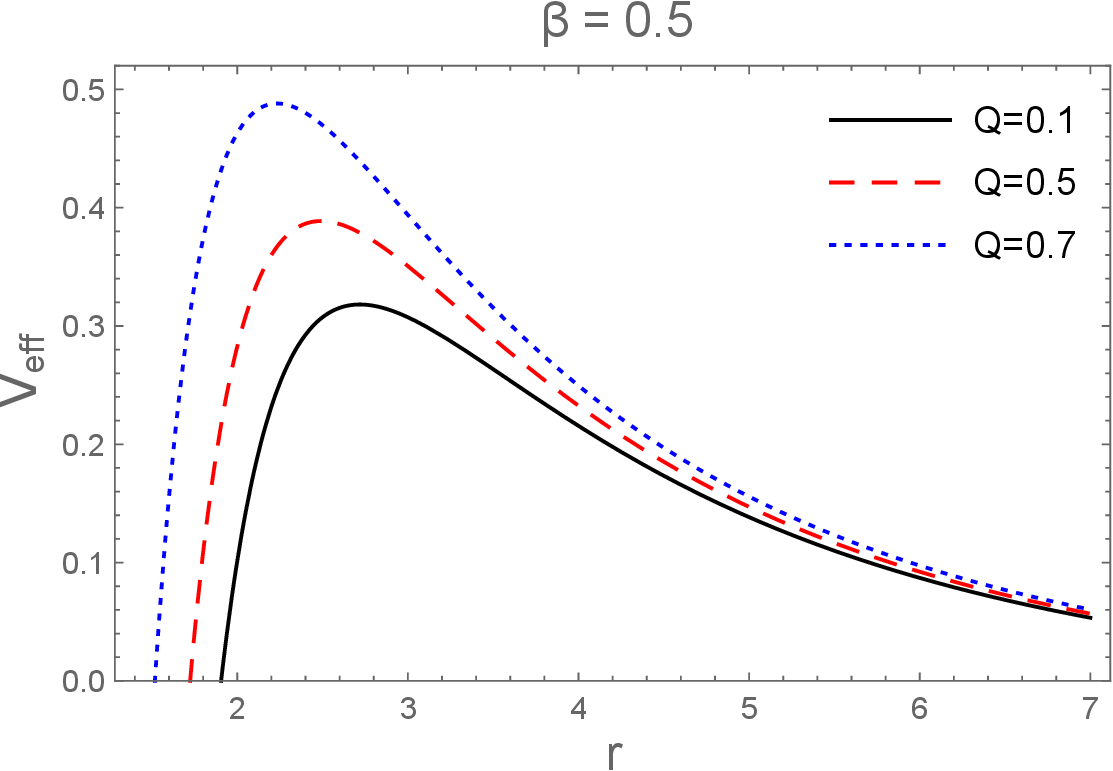}
  \end{center}
\caption{In the upper panel, effective potential is plotted as a function of $r$ for different values of angular momentum with a fixed spin parameter $a=0.5$. In the lower panel, $V_\mathrm{eff}$ is plotted as a function of $r$  for different values of charge parameter with the corresponding fixed parameters, spin $a=0.5$ and angular momentum $L=0.5$.}\label{Veff}
\end{figure*}

\begin{figure*}[t]
 \begin{center}
   \includegraphics[scale=0.4]{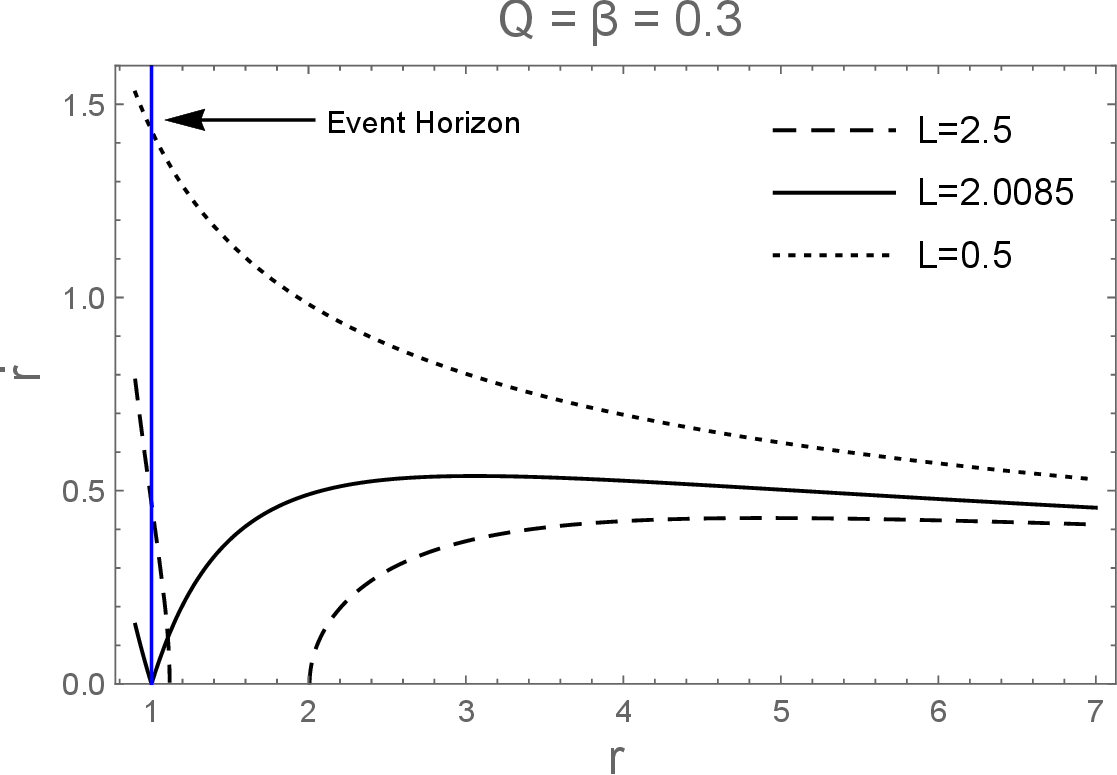}
   \includegraphics[scale=0.4]{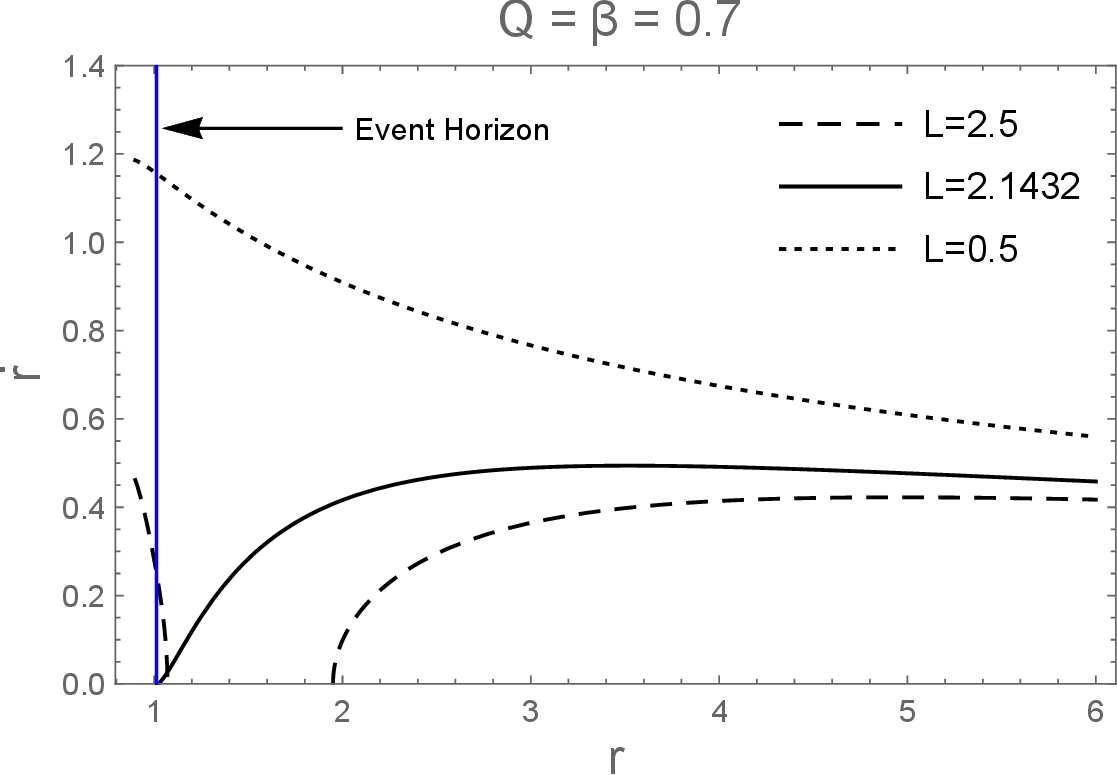}
  \end{center}
\caption{The variation of $\dot{r}$ with respect to the radial coordinate for an extremal black hole.
In the left panel $Q=\beta=0.3$ and $a_E=0.954915061906$ and in the right panel $Q=\beta=0.7$ and $a_E=0.72132723117$.}\label{Radial}
\end{figure*}

\begin{figure*}[t]
 \begin{center}
   \includegraphics[scale=0.5]{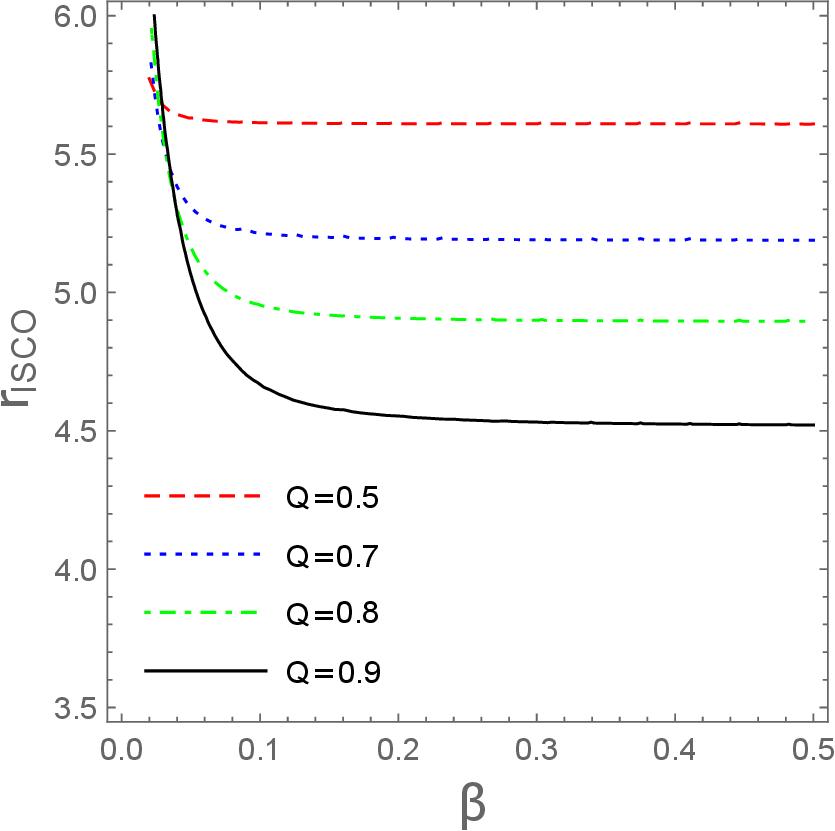}
   \includegraphics[scale=0.5]{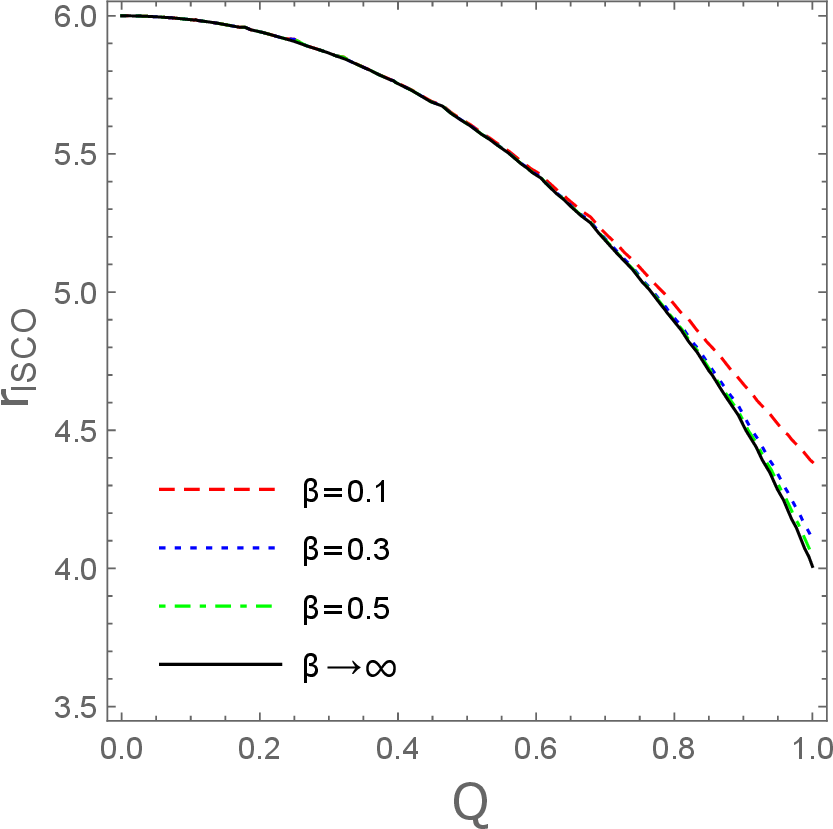}
  \end{center}
\caption{The inner most stable circular orbits by varying the Born-Infeld parameter $\beta$ and
charge $Q$ for a static non-rotating EBI black hole.}\label{Isco}
\end{figure*}

\begin{align}
P_t&=g_{tt}\dot{t}+g_{t\phi}\dot{\phi}, \label{energya}\\
P_{\phi}&=g_{\phi\phi}\dot{\phi}+g_{t\phi}\dot{t}\label{momentum},
\end{align}
\\
where $P_t$ and $P_{\phi}$ are the constants of motion. Basically,
the two quantities $P_t$ and $P_{\phi}$ correspond to the particle’s energy $E$
and the angular momentum $L$, respectively, acting along the axis
of symmetry. The overdot denotes differentiation with respect to the proper time $\tau$. The equations of motion of a massive particle are calculated from (\ref{energya},\ref{momentum})
along with the normalization condition $u_{\mu}u^{\mu}=-m_0^2$, given as below
\\
\begin{align}
\dot{t}&=\frac{1}{r^2}\bigg[\frac{(a^2+r^2)}{\Delta}\big(E(a^2+r^2)-a L\big)
\nonumber \\ &+a\bigg(L-a E\bigg)\bigg], \label{motiona}\\
\dot{\phi}&=\frac{1}{r^2}\bigg[\frac{a}{\Delta}\big(E(a^2+r^2)-a L\big)+\bigg(L-a E\bigg)\bigg],\label{motionb}\\
\dot{r}&=\pm\frac{\sqrt{\big(a L- (a^2 + r^2) E\big)^2 - \Delta\big( m_{0}^2 r^2 + (L - a E)^2\big)}}{r^2}.\label{motionc}
\end{align}
\\
The $+$ and $-$ signs of ~(\ref{motionc}) refer respectively, to the outgoing
and incoming geodesics. In order to understand the motion of the test particle in the vicinity of EBI gravity
thoroughly we must evaluate the effective potential, which is straightforwardly worked out using ~(\ref{motionc}).
\begin{align}
\frac{1}{2}\dot{r}^2&+V_{\mathrm{eff}}=0,\\
V_{\mathrm{eff}}&=-\frac{\big(a L- (a^2 + r^2) E\big)^2 - \Delta\big( m_{0}^2 r^2 + (L - a E)^2\big)}{2r^4}.
\end{align}
\\
When an accelerated particle reaches the black hole, it can most probably continue its motion in the spacetime gravity.
In Fig.~(\ref{Veff}) the effective potential is shown by varying the angular momentum of the incoming test particle for a fixed $Q=\beta$. It is viewed that the potential barrier rises for the larger values of $L$
interpreting that a boosted particle can quickly start circling the black hole.
Also, the feasibility of a particle's motion in the black hole surroundings is increased if the electric field intensity attains its
maximum strength.
\\
The magnitude of a particle's momentum plays a vital role to perceive its geodesics in the
gravitational space-time. Thus one may get the critical value of the angular momentum from (\ref{motiona}) when
$r\rightarrow r^E_H$, i.e, $L_c=\frac{(a^2+(r^E_H)^2)E}{a}$. Fig.~(\ref{Radial}) gives a comprehensible demonstration
of the geodesics in EBI space-time. The particle with $L<L_c$ is always captured by the black hole
gravity and falls exactly at the horizon if $L=L_c$, however, when $L>L_c$ the geodesics never fall into the black hole.
\\
The solution to the simultaneous equations $\partial_{r} V_\mathrm{eff}=\partial^2_{r} V_\mathrm{eff}=0$ defines
the innermost stable circular orbit $r_\mathrm{ISCO}$ of the particle \cite{Shay:2014a,Zasl:2015a,Vrba:2020a,Babar:2016a,Babar:2017a,Shay:2020a,Shay:2020b}.
Fig.~(\ref{Isco}) illustrates the $r_\mathrm{ISCO}$ for a static non-rotating EBI gravity by varying the Born-Infeld parameter
and charge $Q$ of the black hole. It is observed that the radius of the orbits becomes smaller as $\beta$ and $Q$ increases, however the decrease
is seen to be relatively higher in case of the black hole's charge. More precisely we can say that
the charge of the EBI black hole besides its infinite gravity
significantly enhances its ability to grasp the incoming particle
in its vicinity and by increasing the amount of charge the ISCO as a result comes closer to the black hole.
This behaviour is reminiscent of what has been investigated for the Kerr-Newman spacetime in \cite{Liu:2017a}.

\subsection{Near Horizon collision }
Now we analyse the ultrahigh energy produced as a result of a two-particle collision near the horizon
of EBI black hole. We consider particles with the same mass $m_0$ and different four-velocities
$\mathrm{u_1}$ and $\mathrm{u_2}$. The CM energy $E_\mathrm{c.m}$ of collision between two particles at the radial
coordinate $r$ is given by the following expression \cite{Bds:2009a},

\begin{align}
E_\mathrm{c.m}=m_0\sqrt{2}\sqrt{1-g_{\mu\nu}\mathrm{u_{1}}^{\mu}\mathrm{u_{2}}^{\nu}}.\label{Cme}
\end{align}
\\
By substituting (\ref{motiona}-\ref{motionc}) in the above mentioned energy frame we get

\begin{align}
\frac{E_\mathrm{c.m}^2}{2 m{_0}^2 }&=-\frac{\mathcal{K}}{r^2\Delta},
\end{align}
\\
where $\mathcal{K}$ is given as below,

\begin{align}
\mathcal{K}&=-2r^4+2r^3-r^2\big(2a^2+Q^2(r)-L_1L_2\big)
\nonumber \\&-2a^2r+2r\big[a(L_1+L_2)-L_1L_2\big]\nonumber \\ & +Q^2(r)\big(a-L_1\big)\big(a-L_2\big)
\nonumber \\&+\sqrt{\big(a^2+r^2-aL_1\big)^2-\Delta\big[r^2+(a-L_1)^2\big]}\nonumber \\&
\times \sqrt{\big(a^2+r^2-aL_2\big)^2-\Delta\big[r^2+(a-L_2)^2\big]}. \label{energy}
\end{align}

\begin{table}[ht]
\centering
\scalebox{0.9}{\begin{threeparttable}
\caption{\small{The limiting values of the angular momentum for different
extremal cases of a rotating EBI black hole.}}\label{Table3}
 \begin{tabular}{|c c c c c c c|}
 \hline\hline
  Q &  $\beta$ &  $Q(r)_{E}$ & $a_{E}$ &  $r^{E}_{H}$ &  $L_{1}$ & $L_{2}$ \\
 \hline\hline
 0   & 0    &0        &1          &1        &-4.82843 &2\\
 0.2 &0.2   &0.19792  &0.98021707 &1.00141  &-4.80013  &2.00327  \\
 0.3 &0.3   &0.296864 &0.95491506 &1.00304  &-4.76386  &2.0085  \\
 0.4 &0.4   &0.39579  &0.91832659 &1.00512  &-4.71127  &2.01845  \\
 0.5 &0.5   &0.494701 &0.86903103  & 1.00751 &-4.64012  &2.03709   \\
 0.6 &0.6   &0.593595 &0.80470045 & 1.01008 &-4.54673 &2.07259  \\
 0.7 &0.7   &0.692477 &0.72132723 & 1.01274 &-4.42477 &2.1432 \\
 \hline
 \end{tabular}
\end{threeparttable}}
\end{table}

\begin{table}[ht]
\centering
\scalebox{0.9}{\begin{threeparttable}
\caption{\small{The limiting values of the angular momentum for
 different non-extremal cases of a rotating EBI black hole.}}\label{Table4}
\begin{tabular}{|c c c c c c c|}
 \hline\hline
  Q &  $\beta$ &   $a$ &  $r^-_{H}$ & $r^+_{H}$& $L_{1}$ & $L_{2}$ \\
 \hline\hline
 0.2 &0.2   &0.9 & 0.663948  &1.38792 & -4.74224 &2.56572\\
 0.3 &0.3   &0.8 & 0.632149  &1.52037 & -4.64939 &2.78645\\
 0.4 &0.4   &0.7 & 0.638054  &1.59260 & -4.54509 &2.93519\\
 0.5 &0.5   &0.6 & 0.653747  &1.62586 & -4.42765 &3.04241\\
 0.6 &0.6   &0.5 & 0.674623  &1.62645 & -4.29465 &3.11902\\
 0.7 &0.7   &0.3 & 0.686800  &1.65050 & -4.04668 &3.33159\\
 \hline
 \end{tabular}
 \end{threeparttable}}
\end{table}

\begin{figure*}[t]
 \begin{center}
   \includegraphics[scale=0.3]{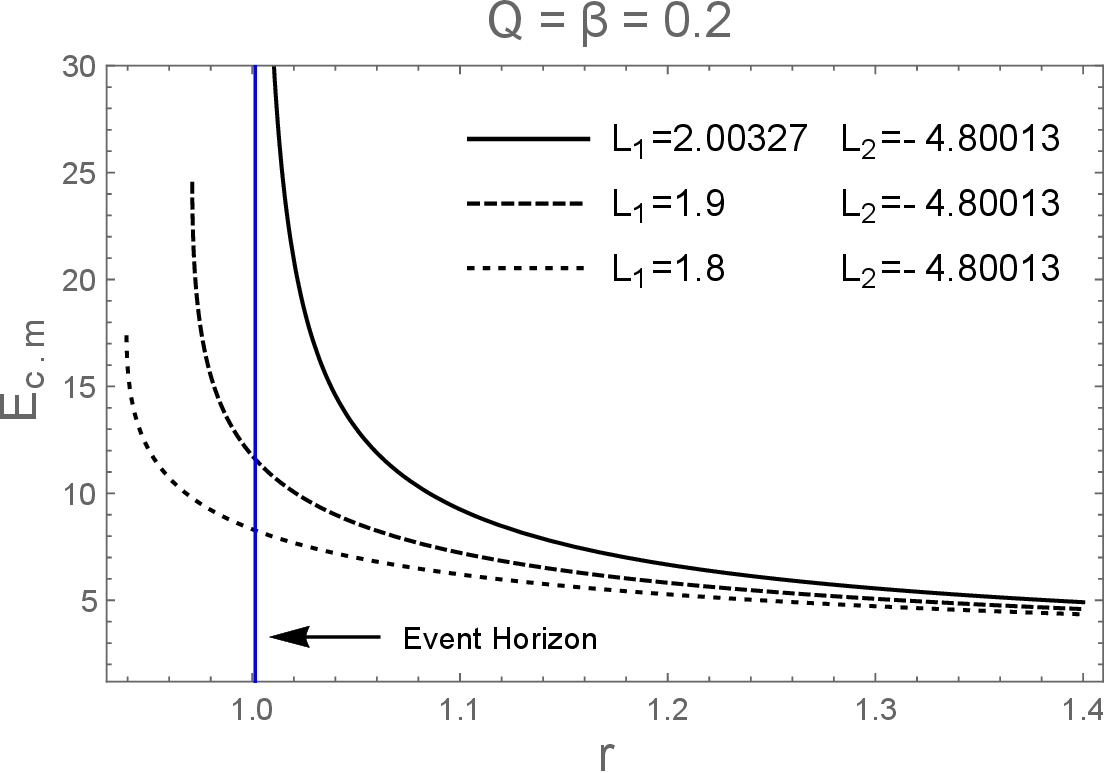}
   \includegraphics[scale=0.3]{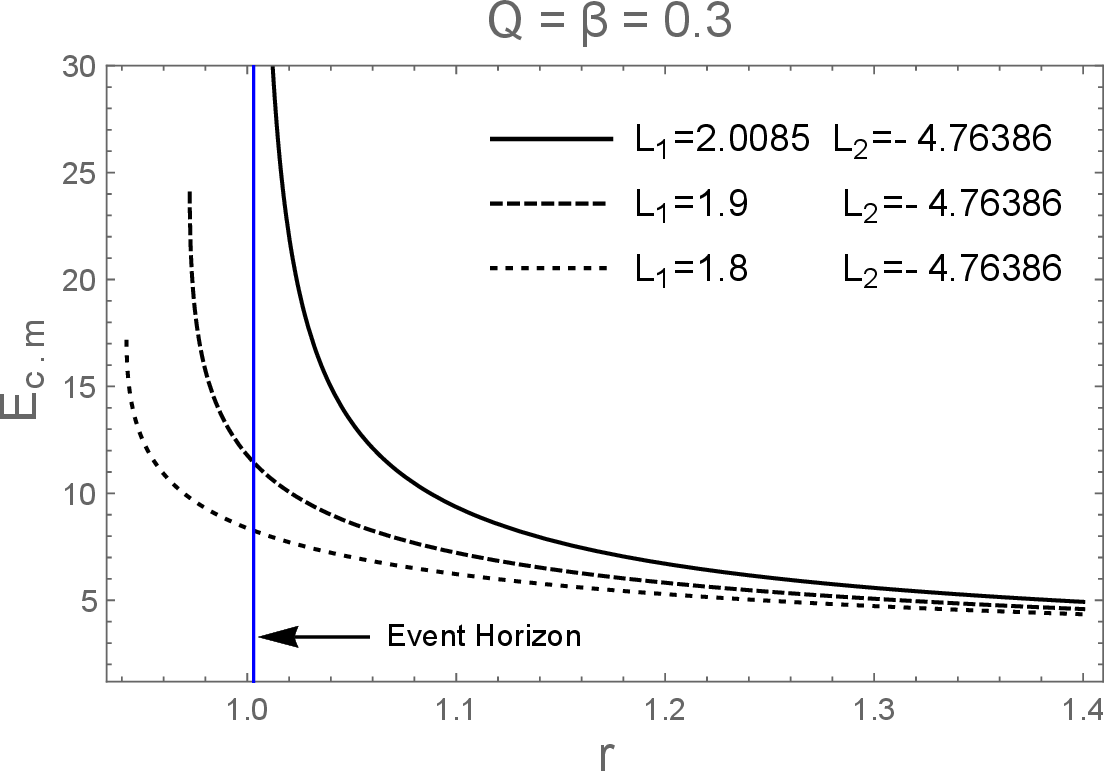}
   \includegraphics[scale=0.3]{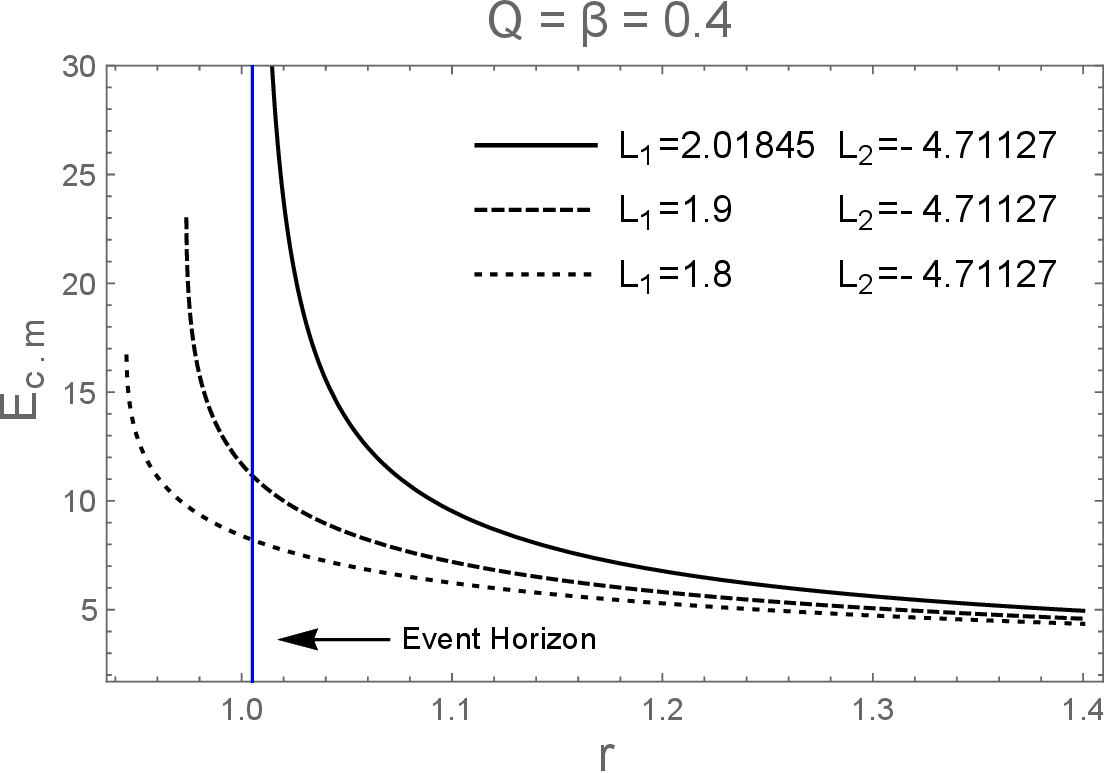}

   \includegraphics[scale=0.3]{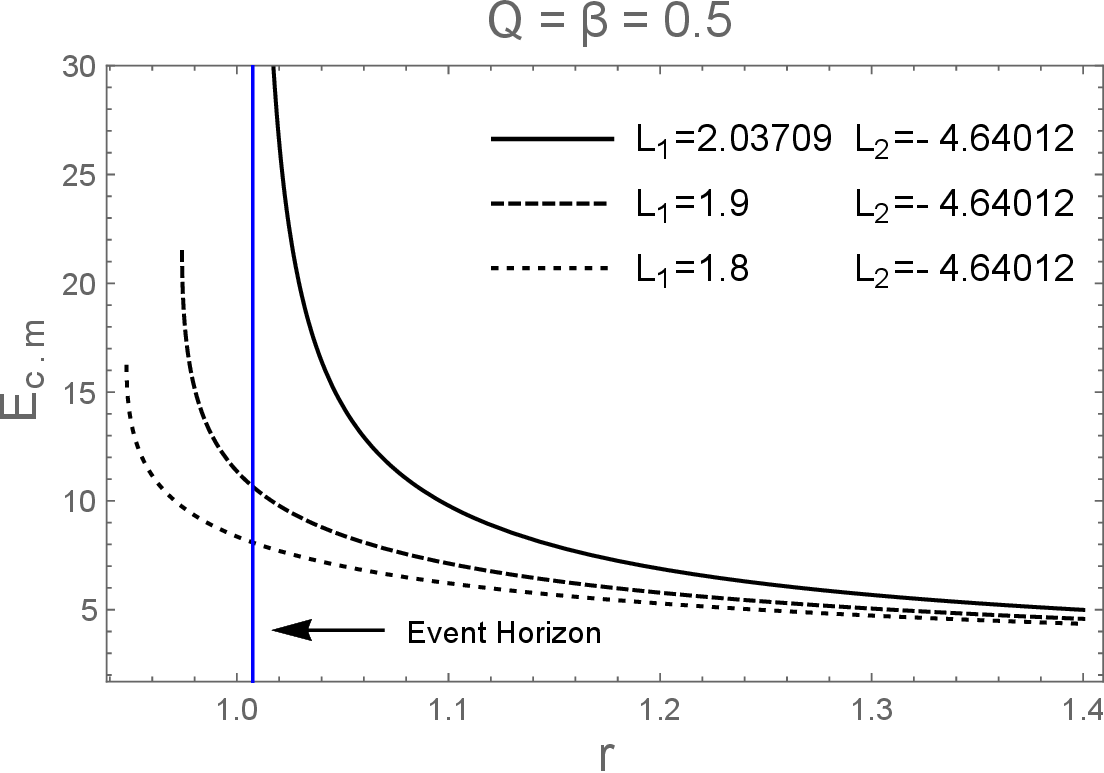}
   \includegraphics[scale=0.3]{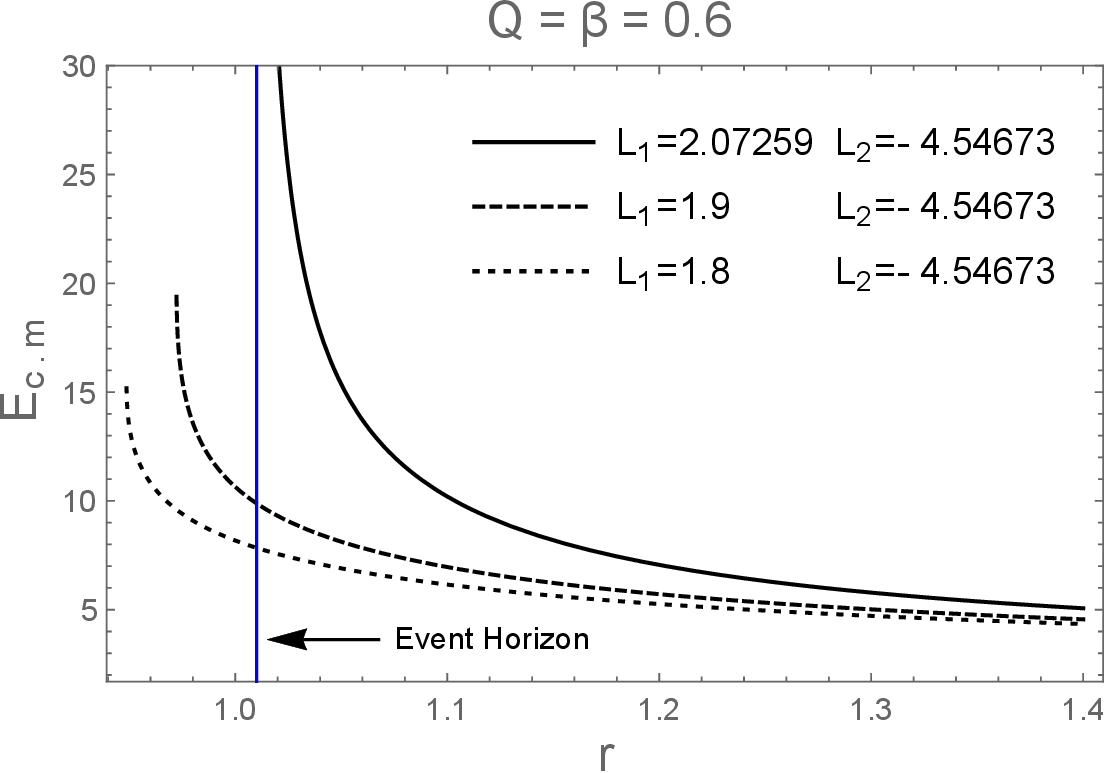}
   \includegraphics[scale=0.3]{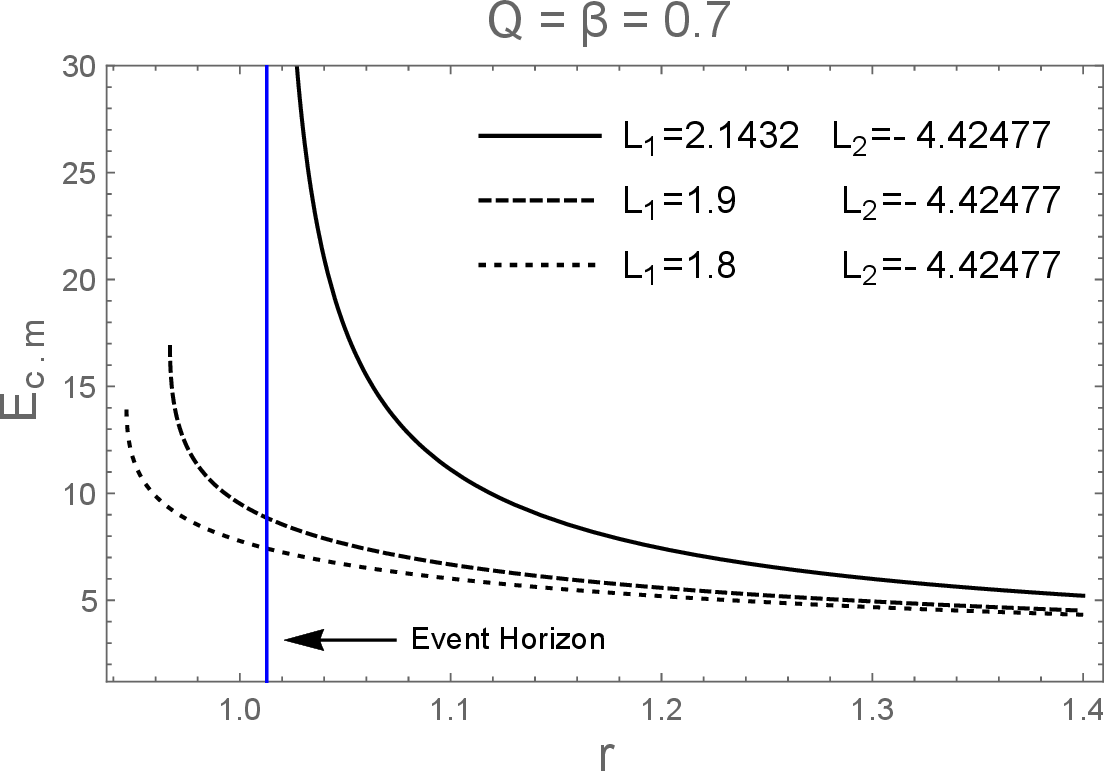}
  \end{center}
\caption{The center-of-mass energy $E_\mathrm{c.m}$ dependence of the radial
 coordinate $r$ for an extremal black hole for various $Q=\beta$ values.}\label{Energyextremal}
\end{figure*}
\begin{figure*}[t]
 \begin{center}
   \includegraphics[scale=0.4]{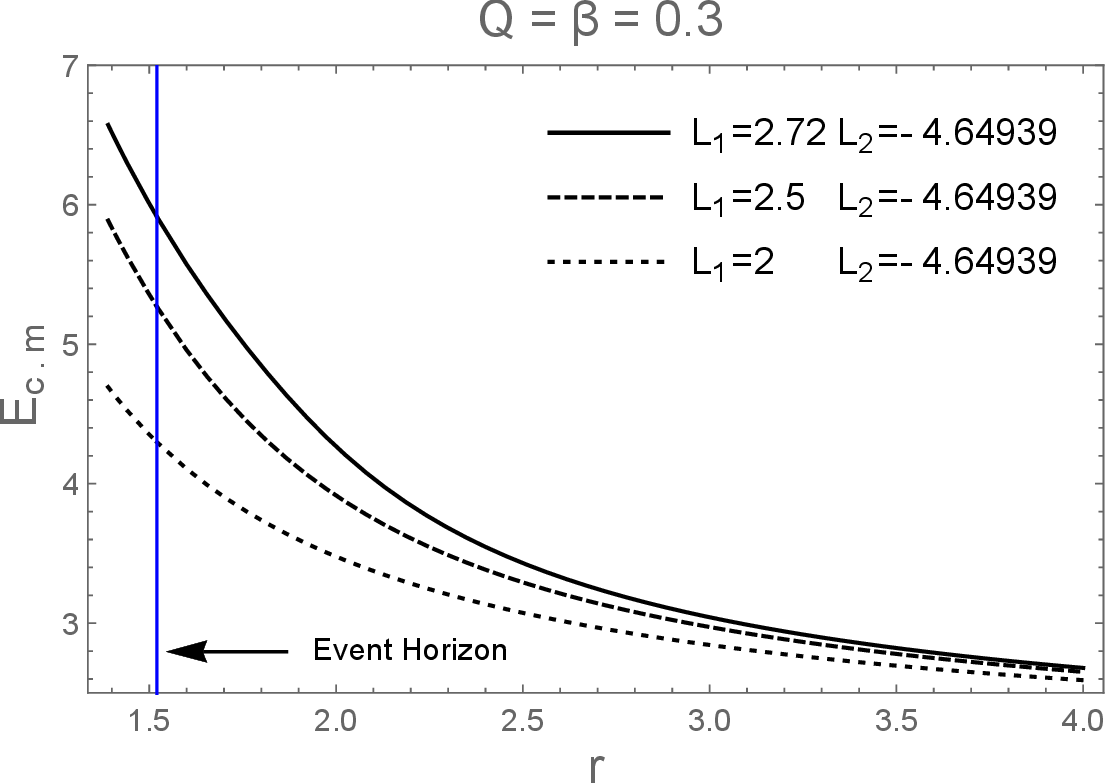}
   \includegraphics[scale=0.4]{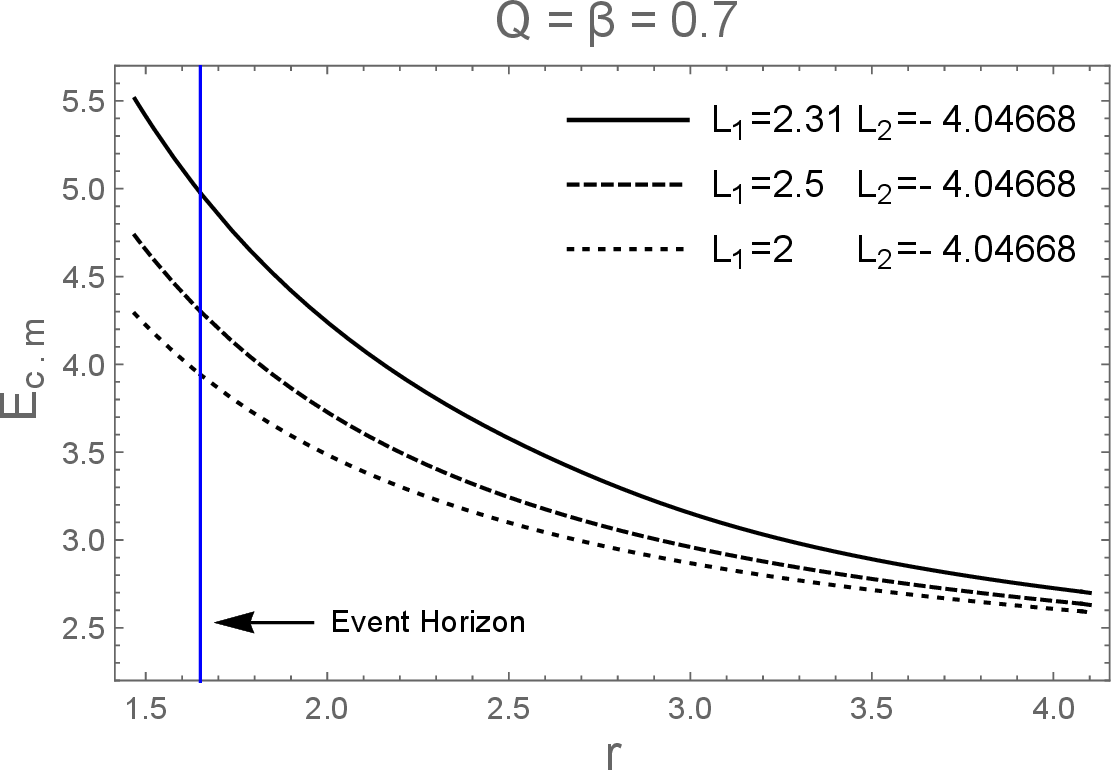}
  \end{center}
\caption{The center-of-mass energy $E_\mathrm{c.m}$ dependence of the radial
 coordinate $r$ for a non-extremal black hole. In the left panel $a=0.8$ and in the right panel $a=0.3$.}\label{Energynonextremal}
\end{figure*}

In our discussion, the participating particles have the same intrinsic identities and are mainly
distinguished by their angular momenta $L_1$ and $L_2$. Here, for the sake of simplicity, we shall take
the conserved energies $E_1/m_0$=$E_2/m_0$=1. It is worth mentioning that an arbitrarily high amount of energy
is obtained when the test particle approaching the back hole has the critical angular momentum $L_c$ \cite{Josh:2015a,Josh:2016a}.
The limiting values of the angular momentum along with the corresponding spin parameters and the horizons
for the \textit{extremal} and \textit{non-extremal} EBI space-time are presented, respectively,
in the Tables (\ref{Table3},\ref{Table4}). The $E_\mathrm{c.m}$ generated as a
result of collision near the horizon of an extremal black hole for different values of $Q=\beta$ is shown in Fig.~(\ref{Energyextremal}).
Quite similar to a charged Kerr-Newman gravity \cite{Wei:2011a}, the CM energy instantaneously diverges near the EBI horizon whenever the incoming particle is equipped with the critical parameters of the motion, on the other hand,
the particles admitting $L<L_c$ contribute only a finite $E_\mathrm{c.m}$. Nonetheless, if we consider a collision
in a non-extremal space-time background, we attain a limited $E_\mathrm{c.m}$ irrespective of the event's location, see Fig.~(\ref{Energynonextremal}).

\section{Strong-field lensing }\label{stronglens}
In this section we shall focus on the strong deflection limit of EBI black hole in the equatorial plane ($\theta=\pi/2$), i.e., both the observer and the source are limited to the equatorial plane. Also, to proceed further we introduce the following dimensionless variables
\\
\begin{eqnarray}
x\to\frac{r}{2M}&,~~t \to \frac{t}{2M},~~a \to \frac{a}{2M},~~q \to \frac{Q}{2M},\nonumber &\\~~\beta \to \frac{\beta}{2M},
\end{eqnarray}
\\
The metric (\ref{metric}) accordingly reduces to the form
\begin{equation}\label{metric1}
ds^2 = -A(x)dt^2 + B(x)dx^2 + C(x)d\phi^2- 2 D(x)dtd\phi,
\end{equation}
\\where
\begin{eqnarray}
A(x) &=& \frac{\Delta-a^2}{x^2}, \\
B(x) &=& \frac{x^2}{\Delta}, \\
C(x) &=& x^2+a^2\left(2-\frac{\Delta-a^2 }{x^2}\right),\\
D(x) &=& a\left(1-\frac{\Delta-a^2 }{x^2}\right),
\end{eqnarray}
\\
and $\Delta=x^2+a^2-x+q^2(x)$.
\begin{eqnarray} \label{qx}
 q^2(x)&\approx\frac{2\beta^2x^4}{3}\bigg(1-\sqrt{1+\xi^2(x)}\bigg)
 \nonumber \\&+\frac{4 q^2}{3}\bigg(1 -\frac{1}{10}\xi^2(x)+\frac{1}{24}\xi^4(x)\bigg).
\end{eqnarray}

\begin{figure*}[t]
 \begin{center}
 \begin{tabular}{c c}
   \includegraphics[scale=0.7]{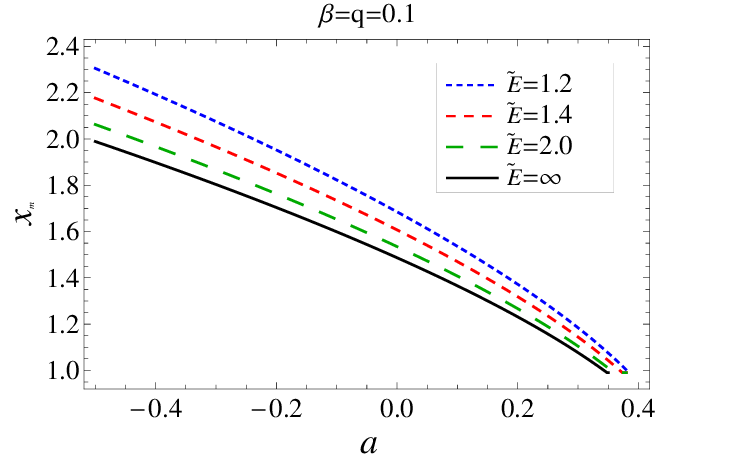}&
   \includegraphics[scale=0.7]{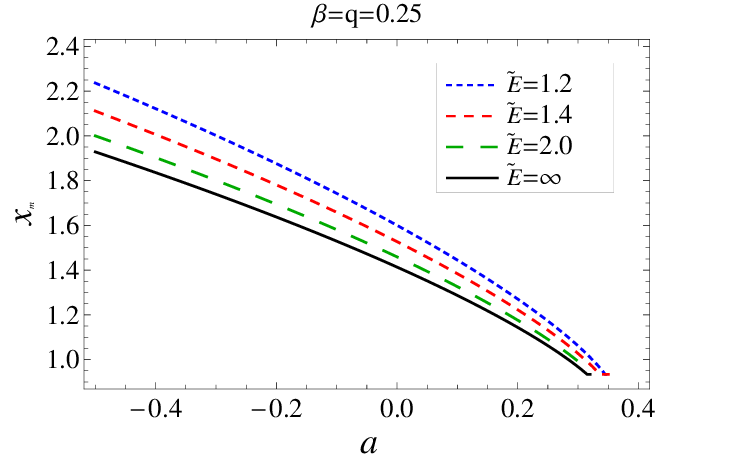}
   \end{tabular}
  \end{center}
\caption{Plot showing the variation of  photon sphere radius $x_m$ with  black hole spin parameter $a$ in different plasma medium. $\tilde{E}=\infty$ corresponds to vacuum. }\label{ps}
\end{figure*}

We assume that the spacetime is filled with a non-magnetized cold plasma whose electron plasma frequency depends on the radius coordinate. We consider the medium to be homogenous but dispersive such that refractive index is constant in space but depends on the wave frequency. The gravitational deflection angle, in this case, is different from the vacuum and depends on the photon frequency. Synge developed the general relativistic geometrical optics in a curved spacetime in a dispersive medium with the reflective index's angular isotropy. In the subsequent analysis we shall take into consideration a static homogeneous plasma in the gravitational field with a refractive index $n$
obtained by Synge \cite{Synge:1960b}

\begin{align}
n^2=1+\frac{p_{\alpha} p^{\alpha}}{(p_{\beta}u^{\beta})^2},
\end{align}
\\
$p_{\alpha}$ and $u^{\beta}$ indicate the four-momentum and four-velocity of the massless particle. One may retrieve
the vacuum case when $n=1$. We define a specific form of the refractive index $n$ for a precise analysis \cite{Bin:2010a,Tsp:2011a}

\begin{align}
n^2=1-\frac{{\omega^2_{e}}}{[\omega(x^i)]^2}.\label{refractiveindex}
\end{align}
\\
Here, $\omega_e$ is the plasma electron frequency and $\omega(x^i)$ is
the photon frequency depending on the spatial coordinates $x^i$  detected by a distant observer.
The light can propagate through the plasma medium if and only if the inequality, $\omega^2>{\omega^2_{e}}$ is satisfied.
The plasma frequency admits the following analytic expression

\begin{align}
 \omega^2_{e}&=\frac{4\pi e^2 N(x^i)}{m}=K_{e}N(x^i),\label{omegae}
 \end{align}
\\
where $e$, $m$ and $N(x^i)$ are the charge, mass and number density of the electron, respectively \cite{Rog:2015a}. The Hamiltonian for the photon around the EBI black hole surrounded by plasma has the following form
\begin{align}\label{hamilton}
H(x^i,p_i) = \frac{1}{2}\left( g^{ik}p_i p_k +\omega_e^2 \right)=0,
\end{align}
\\
where $g^{ij}$ is the contravariant metric tensor. Using Eq.(\ref{hamilton}), the Hamiltonian differential equations which are given by
\begin{align}
\frac{dx^{i}}{d\lambda}=\frac{\partial H }{\partial p_i}, ~~~~~~\frac{dp_{i}}{d\lambda}=-\frac{\partial H }{\partial x^i},
\end{align}
\\
give two constants of motion which are Energy $E$ and
angular momentum $L$ of the photon. The conserved quantities $E$ and $L$, attain the form
\begin{eqnarray}
E=-p_t=\omega ~~~~~~~ L=p_{\phi}.
\end{eqnarray}
\\
We introduce notation $\tilde{L}$ and $\tilde{E}$, by considering a homogeneous plasma with $\omega_e = constant$, as
\begin{equation}
\tilde{E}=\frac{-p_t}{\omega_e}=\frac{\omega}{\omega_e} ~~~~~ \tilde{L}=\frac{p_{\phi}}{\omega_{e}}.
\end{equation}
\\
The first order geodesic equation in the equatorial plane in terms of $\tilde{E}$ and $\tilde{L}$ reduce to the following form

\begin{eqnarray}\label{tdot}
\dot{t} &=& \frac{\omega_e\left(\tilde{E}C(x)-\tilde{L}D(x)  \right)}{A(x)C(x)+D(x)^2}, \nonumber\\
&=& \frac{\omega_e}{x^2}\bigg[\frac{(a^2+x^2)}{\Delta}\big(\tilde{E}(a^2+x^2)-a \tilde{L}\big)+a\big(\tilde{L}-a \tilde{E}\big)\bigg],
\end{eqnarray}

\begin{eqnarray}\label{pdot}
\dot{\phi} &=& \frac{\omega_e\left(\tilde{E}D(x)+\tilde{L}A(x)  \right)}{A(x)C(x)+D(x)^2}, \nonumber\\
&=& \frac{\omega_e}{x^2}\bigg[\frac{a}{\Delta}\big(\tilde{E}(a^2+x^2)-a \tilde{L}\big)+\big(\tilde{L}-a \tilde{E}\big)\bigg],
\end{eqnarray}

\begin{align}\label{rdot}
\dot{x}^2 &=\frac{\omega_e^2}{B(x)\big(A(x)C(x)+D(x)^2\big)} \bigg(\tilde{E}^2 C(x)-2 \tilde{L}\tilde{E}D(x)\nonumber \\ &-\tilde{L}^2 A(x)-
\big(A(x)C(x)+D(x)^2\big) \bigg), \nonumber\\
&=\frac{\omega_e ^2}{x^4}\bigg[\big(a \tilde{L}- (a^2 + x^2) \tilde{E}\big)^2 - \Delta\big(x^2 + (\tilde{L} - a \tilde{E})^2\big)\bigg].
\end{align}
\\
With the condition $\dot{x}\Big|_{x=x_0}=0$, Eq.~(\ref{rdot}) can be solved for $\tilde{L}$ to give
\begin{eqnarray}
\tilde{L(x_0)} &=& \frac{-\tilde{E}D(x_0) + \sqrt{A(x_0) G(x_0)+\tilde{E}^2 D(x_0)^2}}{A(x_0)}\\
G(x_0) &=& \tilde{E}^2C(x_0)-(A(x_0)C(x_0)+D(x_0)^2),
\end{eqnarray}
\\
which solves to
\begin{equation}
\tilde{L(x_0)} = \frac{a \tilde{E}(a^2 +  x_0^2 -  \Delta) -x_0 \sqrt{(a^2 - \Delta + \tilde{E}^2 x^2 )\Delta}}{a^2-\Delta}
\end{equation}

The photon traveling close to the black hole encounters a turning points at $\dot{x}=0$, and at radius $x_0 = x_m$, the deflection angle becomes infinitely large. The radius $x_m$ is the photon sphere radius and is the largest solution of the equation
\begin{eqnarray}
\left(G'(x_0) A(x_0) -A'(x_0) G(x_0)\right)\nonumber\\+2 \tilde{L}\tilde{E}\left(A'(x_0) D(x_0)-A(x_0) D'(x_0) \right)
\end{eqnarray}
\\
In Fig.~(\ref{ps}), we present the variation of photon sphere with the black hole spin parameter in vacuum and in different plasma medium. The presence of the medium can increase the photon sphere radius. In the case of $a>0$ (prograde orbit), the photons tend to come closer to the black hole compared to $a<0$ (retrograde orbit). Especially for the extreme values of $a$ , the photon sphere radius in the plasma medium tends to be same as in the vacuum for prograde orbits.

\begin{figure*}[t]
\begin{center}
            \includegraphics[scale=0.65]{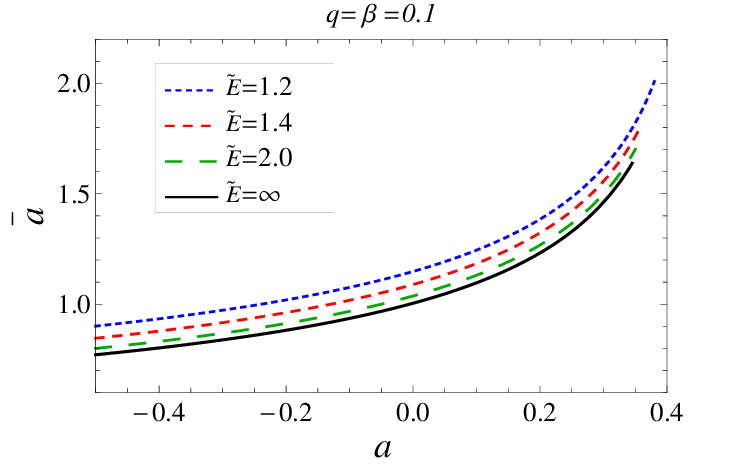}
			\includegraphics[scale=0.65]{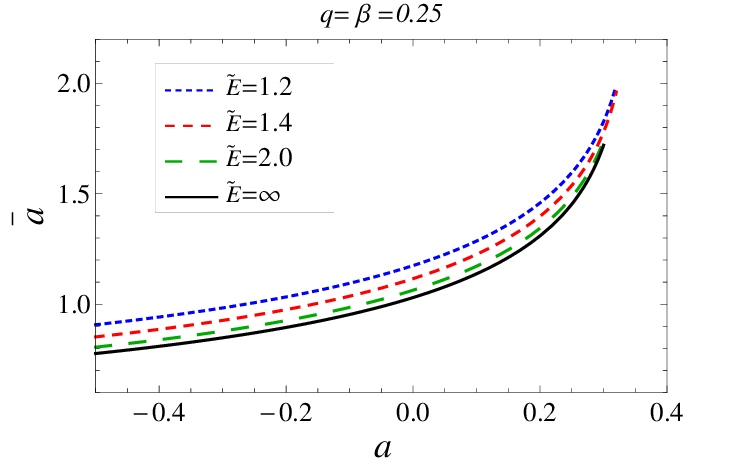}
            \includegraphics[scale=0.65]{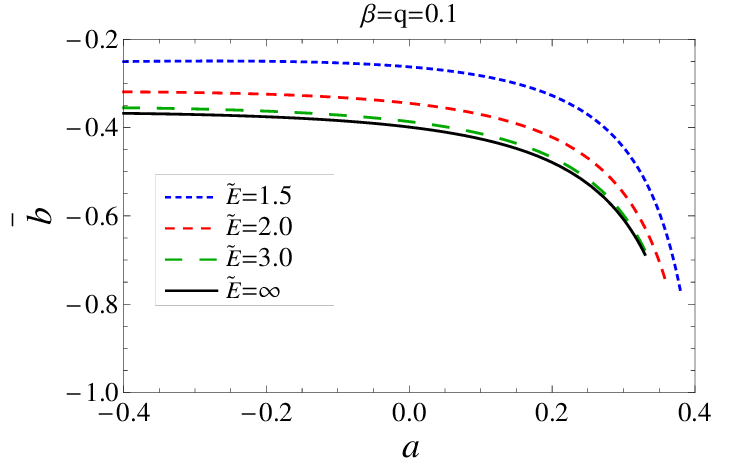}
			\includegraphics[scale=0.65]{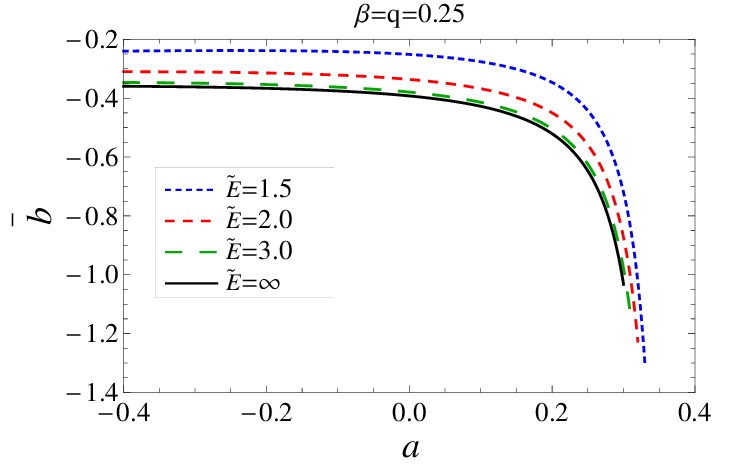}
		    \includegraphics[scale=0.65]{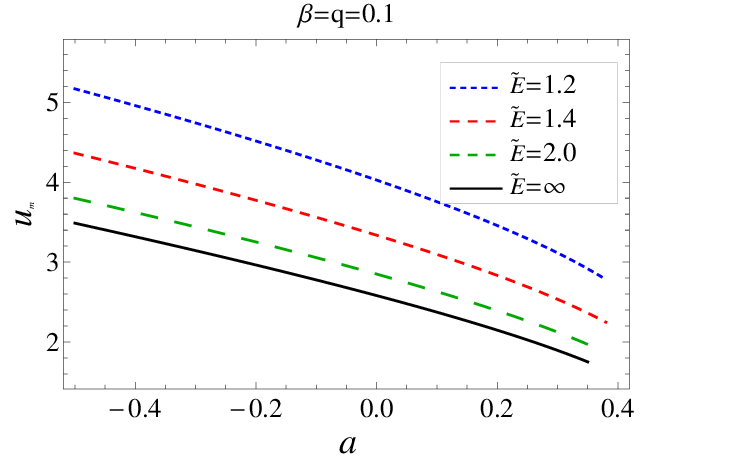}
		    \includegraphics[scale=0.65]{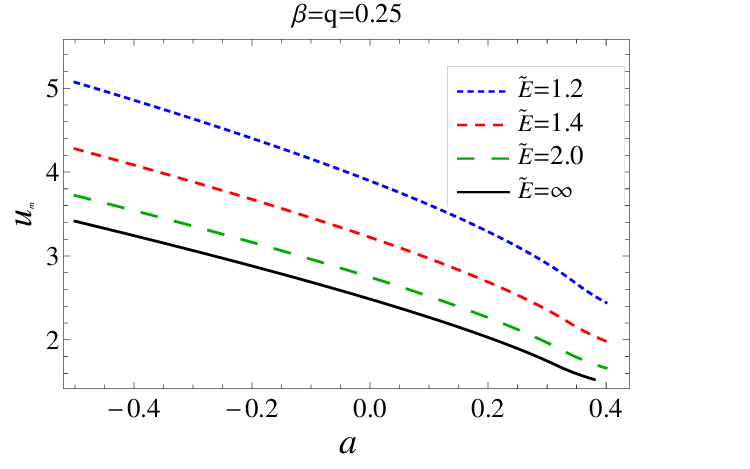}
  \end{center}
	\caption{Plot showing the behavior of strong lensing coefficients $\bar{a}$, $\bar{b}$ and $u_m$ as a functions of $a$, $q$ for different values of  $\tilde{E}$.}\label{plot3}		
\end{figure*}

Bozza \cite{Bozza:2002b,Bozza:2002af} developed the method to find the deflection angle suffered by the light on its way from the source to observer and is given by
\begin{eqnarray}
\alpha_D(x_0) =  I_{T}(x_0)-\pi,\end{eqnarray}
\\
Here, $I_{T}(x_0)$ is the total azimuthal angle, which on using Eq.~(\ref{pdot}) and Eq.~(\ref{rdot}) takes the form
\begin{eqnarray}\label{def3}
I_{T}(x_0) &=& \frac{\sqrt{B(x)A(x_0)}(\tilde{E}D(x)+\tilde{L}A(x))}{\sqrt{D(x)^2+A(x)C(x}\sqrt{P(x,x_0)}},
\end{eqnarray}
\begin{eqnarray}
P(x,x_0)&&= G(x)A(x_0)-G(x_0)A(x)+2\tilde{E}\tilde{L}(A(x)D(x_0)\nonumber\\&& -A(x_0)D(x)).
\end{eqnarray}

\begin{figure*}[t]
  \begin{center}	
   \begin{tabular}{c c c c}
     \includegraphics[scale=0.7]{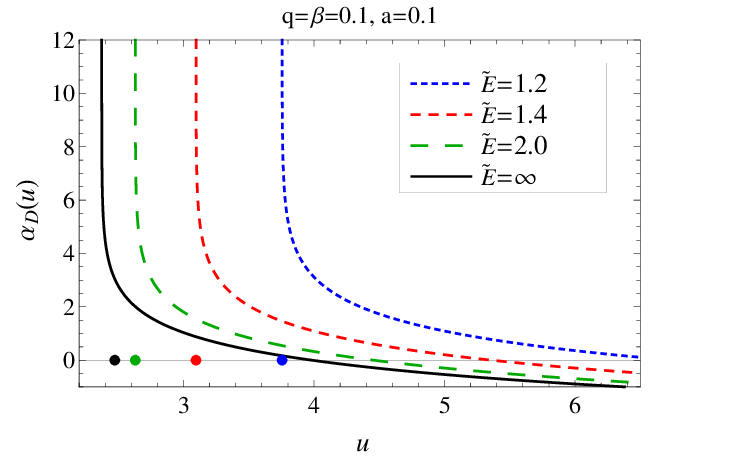}&
     \includegraphics[scale=0.7]{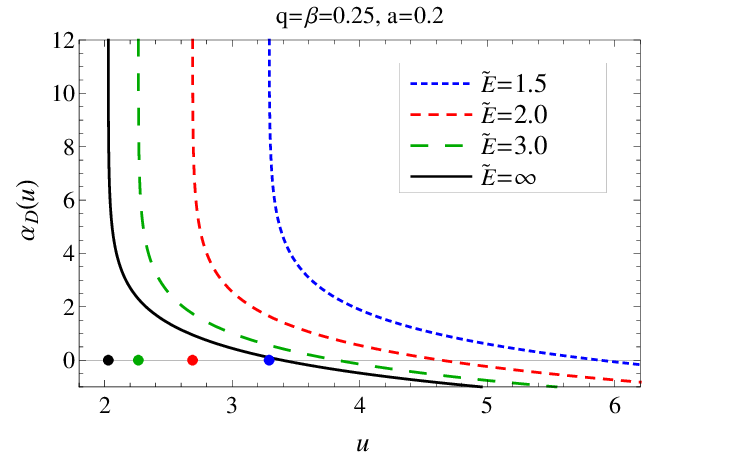}\\
   \end{tabular}
 \caption{Plot showing the behaviour of deflection angle with  the impact parameter for different values of  $a$ , $q$ and $\beta$ for different values of $\tilde{E}$. The points on the horizontal line correspond to the impact parameter at which deflection angle diverges. }\label{plot4}	
   \end{center}
\end{figure*}

As the photon approaches the black hole, $x$ decreases from infinity to $x_0$ and increases again from $x_0$ to infinity as it recedes away. Due to symmetries in space-time, the total change in the azimuthal angle is just twice the change from $x_0$ to infinity. If the black hole were not present the total change in the azimuthal angle is $\pi$. Thus deflection angle is accordingly zero. The deflection angle increases as $x_0$ approaches $x_m$ and diverges logarithmically at $x_0 = x_m$.

Following the method developed by Bozza \cite{Bozza:2002b}, we can have the deflection angle in SDL as
\begin{eqnarray}\label{def4}
\alpha_{D}(u) &=& \bar{a} \log\left(\frac{u}{u_m} -1\right) + \bar{b} + \mathcal{O}(u-u_m).
\end{eqnarray}
The details of the calculation can be found in Appendix A. The lensing coefficients $\bar{a}$ and $\bar{b}$ are smaller
in vacuum than in the plasma, both decreasing with a just
like the Kerr black hole (see Fig. \ref{plot3}). The critical impact
parameter, i.e., where the deflection angle diverges, for a
fixed values of parameters increases as the frequency of the
plasma increases (see Fig. \ref{plot4}).

\subsection{Observables}\label{observable}
A geometrical relationship called the lens equation is set up between the observer, the source, and the lens to discuss the strong gravitational lensing by a black hole. In \cite{Frittelli,Frittelli2,Perlick} there is no requirement for flat background or any condition on the positioning of observer or source, however, to establish a better connection between the models and the observations to get the more clear physical picture we will assume some approximations in the lens equation. The most important and reasonable is the asymptotic approximation i.e., both observer and the source should be in flat spacetime and are not affected by the lens' curvature effects. It allows us to take all the distance measurements using Euclidean geometry. The asymptotic approximation also considers that the source and observer should be far enough from the lens, and for further implication, the source should lie behind the lens. Even cases of source lying at the back of observer or between observer and lens, which is called retrolensing have been studied
\\
\begin{figure*}[t]
	\begin{centering}
		\begin{tabular}{c c}
		    \includegraphics[scale=0.75]{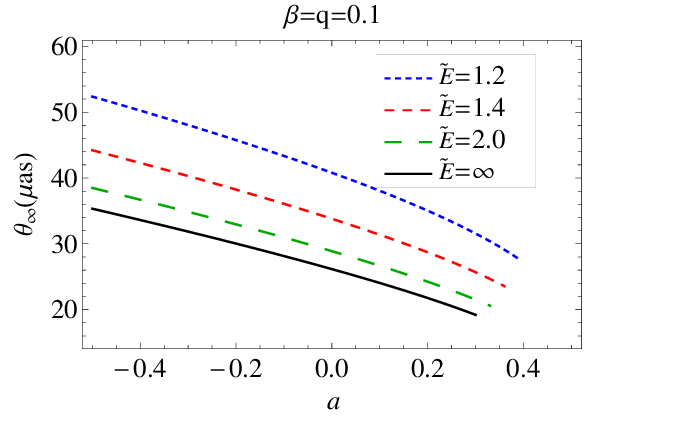}&
			\includegraphics[scale=0.75]{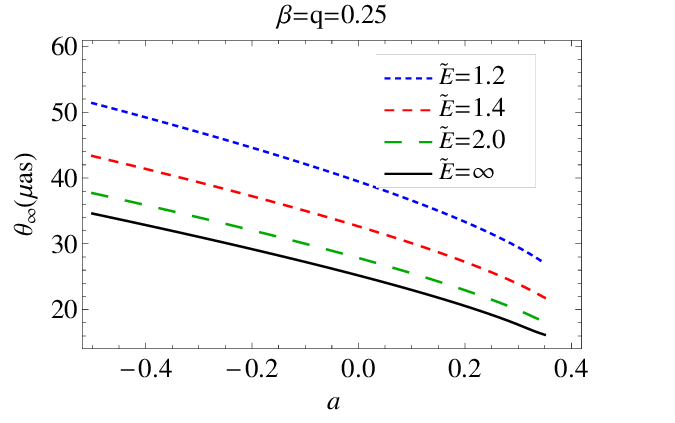}\\
			\includegraphics[scale=0.75]{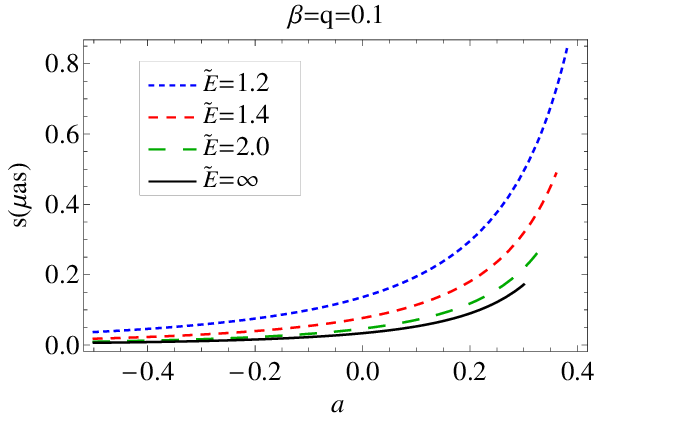}&
			\includegraphics[scale=0.75]{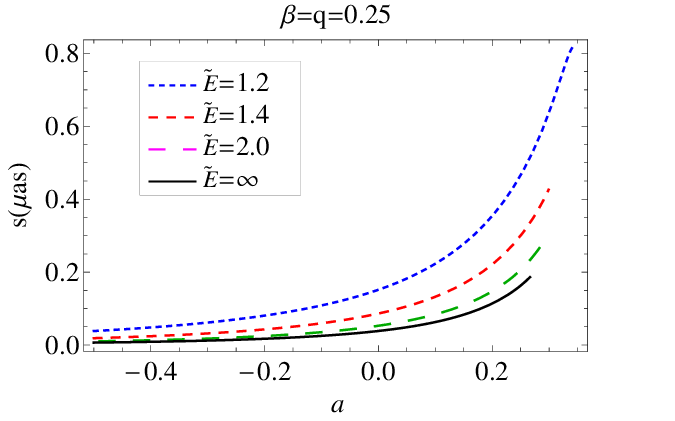}\\
		\end{tabular}
	\end{centering}
	\caption{Plot showing the behavior of strong lensing observables as a function of $a$ , for different values of $q$, $\beta$ and $\tilde{E}$ for SgrA*.}\label{plot5}		
\end{figure*}

\begin{figure*}[t]
	\begin{centering}
		\begin{tabular}{c c}
		    \includegraphics[scale=0.75]{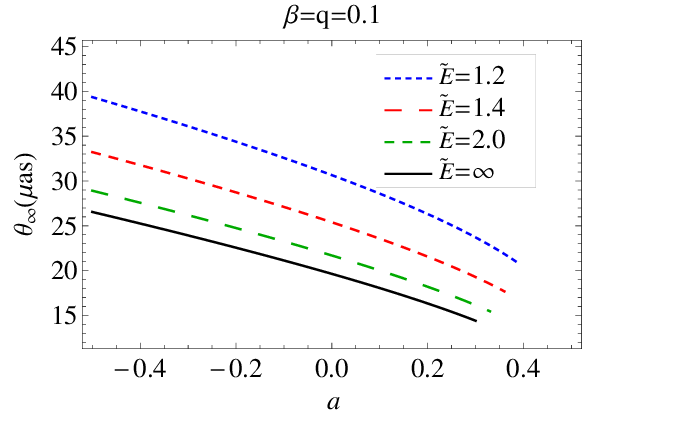}&
			\includegraphics[scale=0.75]{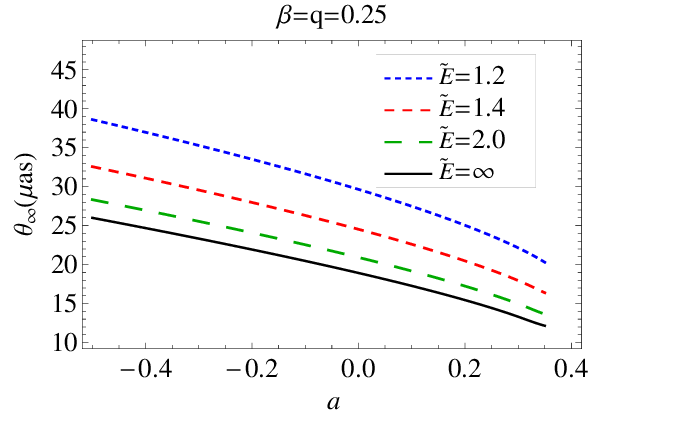}\\
			\includegraphics[scale=0.75]{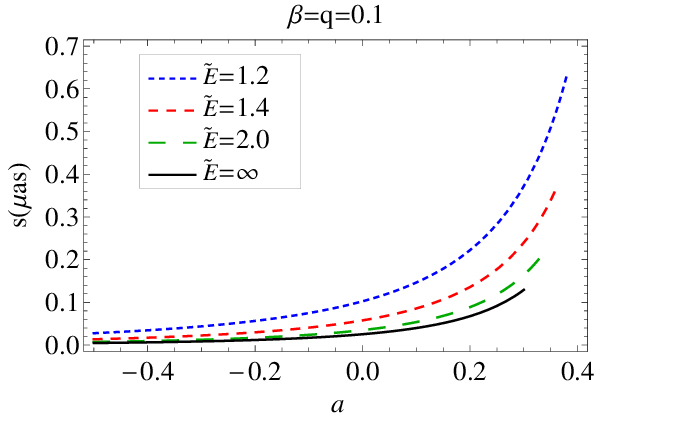}&
			\includegraphics[scale=0.75]{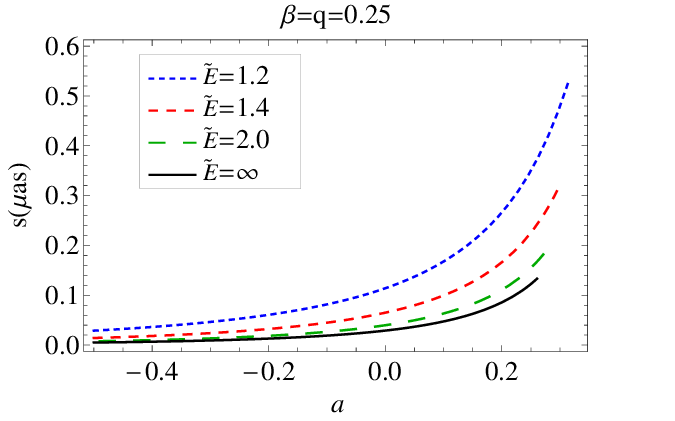}\\
		\end{tabular}
	\end{centering}
	\caption{Plot showing the behavior of strong lensing observables as a function of $a$ , for different values of $q$, $\beta$ and $\tilde{E}$ for M87*.}\label{plot6}		
\end{figure*}

The geometrical configuration for gravitational lensing constitutes a light source $S$, an observer $O$ and a black hole $L$ that acts as a lens and lies between the source and observer. The photons emitted by the source is deviated by the black hole and reach the observer. The observer sees the source's image at an angle $\theta$ from the optical axis $OL$, whereas, the source is at an angular position of $\beta$ from $OL$. The emission direction and the detection direction make an angle $\alpha_{D}(\theta)$, which is called the deflection angle. The closest approach distance $x_0$ is different than the impact parameter $u$ as long as the deflection angle is not vanishing.
\\
Several lens equations, which principally differ from each other for different choices of variables, were introduced however in \cite{Bozza:2008ev}, the author gave a detailed comparison of varying lens equation and suggested that Ohanian lens is the best approximate lens equation. Using the coordinate independent Ohanian lens equation \cite{Oho:1987ev} which connects the source and observer positions as

\begin{eqnarray}\label{oho}
\xi &=& \frac{D_{OL}+D_{LS}}{D_{LS}}\theta-\alpha_{D}(\theta),
\end{eqnarray}
$\xi$ is the angle between the lens-source direction and the optical axis. $D_{OL}$ is the observer-lens distance and $D_{LS}$ is the lens-source distance. The angle $\xi$ and $\beta$ are related by
\begin{eqnarray}\label{rel}
\frac{D_{OL}}{\sin(\xi-\beta)} &=& \frac{D_{LS}}{\sin \beta}.
\end{eqnarray}
The lensing effects are most evident when all the objects are almost aligned. It is the case when the relativistic images are most prominent. So we study the case when the angles $ \beta $, $\xi$ and $\theta$ are minimal. However, if a ray of light emitted by the source $S$ follow multiple loops around the black hole before reaching the observer, $\alpha $ must be very close to a multiple of 2$\pi$. Replacing $\alpha_{D}(\theta)$ by $\alpha_{D}(\theta)-2n\pi=\Delta\alpha _n$, with  $n \in N $ and $ 0 < \Delta\alpha _n \ll 1 $, we can rewrite Eq.~(\ref{oho}) for small values of  $\beta$, $\xi$ and $\theta$ using Eq.~(\ref{rel}) as,
\begin{eqnarray}\label{lensequation}
\beta &=& \theta -\frac{D_{LS}}{D_{OL}+D_{LS}} \Delta\alpha _n.
\end{eqnarray}
Given the angular position of source $\beta$ and the distances of observer and source from the black hole, one can calculate the image positions using Eq.~(\ref{lensequation}). The deflection angle increases as the light ray trajectory get closer to the event horizon, such that for a particular value of $u$, the light form loops around the black hole resulting in $ \alpha_{D}(\theta)> 2\pi$. With further decreasing impact parameter, the light ray winds several times around the black hole before escaping to the observer. Finally, for critical impact parameter, corresponding to the closest distance $x_m$, the deflection angle diverges. For each loop of the light geodesic, there is a particular value of impact parameter at which light reaches the observer from the source. So infinite sequence of images will be formed on each side of the lens. Equation~(\ref{def4}) with $\alpha_{D}(\theta_n{^0}) = 2n\pi $ leads to
\begin{eqnarray}\label{theta}
\theta_n{^0} &=& \frac{u_m}{D_{OL}}(1+e_n),
\end{eqnarray}
where
\begin{eqnarray}
e_n &=& e^{\frac{q-2n\pi}{p}}.
\end{eqnarray}
\\
Here, $n$ is the number of loops followed by the photons around the black hole. The deflection angle is expanded around $\theta_n{^0}$ as:
\begin{equation}
\alpha_{D}(\theta) = \alpha_{D}(\theta_n {^0}) +\frac{\partial \alpha_{D}(\theta)}{\partial \theta } \Bigg |_{\theta_n{^0}}(\theta-\theta_n{^0})+\mathcal{O}(\theta-\theta_n{^0}).
\end{equation}
On using (\ref{theta}) and setting $\Delta\theta_n= (\theta-\theta_n{^0}) $ we obtain
\begin{eqnarray}
\Delta\alpha_n &=& -\frac{\bar{a} D_{OL}}{u_m e_n}\Delta\theta_n.
\end{eqnarray}
Finally the lens equation (\ref{lensequation}) becomes
\begin{eqnarray}
\beta &=& ( \theta_n{^0} + \Delta\theta_n )+\frac{D_{LS}}{D_{OL}+D_{LS}}\Bigg(\frac{\bar{a} D_{OL}}{u_m e_n}\Delta\theta_n\Bigg).
\end{eqnarray}
The second term in the above equation is very small as compared to the third term, there by neglecting it, we get
\begin{eqnarray}\label{angpos}
\theta_n &=& \theta_n{^0} + \frac{D_{OL}+D_{LS}}{D_{LS}}\left(\frac{u_me_n}{\bar{a} D_{OL}}(\beta-\theta_n{^0})\right).
\end{eqnarray}
The above equation, for $\beta>0$, determines images only on the same side of the source ($\theta>0$). To obtain the images on the opposite side, one can solve the same equation with the source placed at $-\beta$. Finally using Eq.~(\ref{angpos}), we obtain full set of primary ($+\beta$) and secondary images ($-\beta$).

If the outermost image can be separated from the inner packed one, we can have following observables as 
\begin{eqnarray}
\theta_\infty &=& \frac{u_m}{D_{OL}},\\
s &=& \theta_1-\theta_\infty \approx  \theta_\infty (e^{\frac{\bar{b}-2\pi}{\bar{a}}}),
\end{eqnarray}

\begin{table*}[t]
\caption{Estimates for lensing observables and strong lensing coefficients for the black hole Sgr A* and M87  for different values of $a$, $q$ and $\tilde{E}$. $R_s = 2GM/c^2$ is the Schwarzschild radius.
\label{table1}
	}
\resizebox{1\textwidth}{!}{

	\begin{tabular}{p{1cm} p{1cm} p{1cm} p{1.2cm} p{1.5cm} p{2.0cm} p{1.2cm} p{2.0cm} p{1cm} p{1.5cm} p{1cm}}
		
\hline\hline
\multicolumn{4}{c}{}&
 \multicolumn{2}{c}{Sgr A*} & \multicolumn{2}{c}{M87*}  & \multicolumn{2}{c}{Lensing Coefficients}\\
{$a$ } & {$q$} & {$\beta$} & {$\tilde{E}$} & {$\theta_\infty $ ($\mu$as)} & {$s$ ($\mu$as) } &    {$\theta_\infty $ ($\mu$as)} & {$s$ ($\mu$as) } &   {$\bar{a}$}&{$\bar{b}$} & {$u_m/R_s$}\\ \hline
0.0   & 0.0  & 0.0 & $\infty$ & 26.3299 & 0.0329518 & 19.782 & 0.0247572 & 1.0000 & -0.40023 & 2.59808  \\
\hline
\multirow{4}{*}{0.0}   & 0.10  & 0.10 & 1.2 & 40.8014 & 0.136406 & 30.6547 & 0.102484 & 1.14818 & -0.262399 & 4.02604  \\
                      & 0.10  & 0.10 & 1.4 & 33.8068 & 0.0768462 & 25.3996 & 0.0577357 & 1.08896 & -0.344909 & 3.33586  \\
                      & 0.10  & 0.10 & 2.0 & 28.8801 & 0.0465492 &21.6981 & 0.0349731  & 1.0372 & -0.386451 & 2.84972  \\
                      & 0.10  & 0.10 & $\infty$ & 26.1546 & 0.033707 & 19.6504 & 0.0253245 & 1.0042 & -0.398845 & 2.58079  \\
\hline
\multirow{4}{*}{0.0}   & 0.25  & 0.25 & 1.2 & 39.4685 & 0.151685 & 29.6533 & 0.113963  & 1.17496 & -0.251322 & 3.89452\\
                      & 0.25  & 0.25 & 1.4 & 32.6493 & 0.0864492 & 24.5299 & 0.0649506 & 1.11551 & -0.336246 & 3.22164\\
                      & 0.25  & 0.25 & 2.0 & 27.8451 & 0.052903 & 20.9204 & 0.0397468  &1.06329 & -0.379309 & 2.74759\\
                      & 0.25  & 0.25 & $\infty$ & 25.1872 & 0.0385673 & 18.9235 & 0.0289762 & 1.02991 & -0.392364 & 2.48532\\
\hline
\multirow{4}{*}{0.1}   & 0.10  & 0.10 & 1.2 & 38.0561 & 0.194843 & 28.5922 & 0.146388  & 1.24476 & -0.282442 & 3.75516\\
                      & 0.10  & 0.10 & 1.4 & 31.3725 & 0.114023 & 23.5706 & 0.0856672 & 1.18444 & -0.37017 & 3.09565\\
                      & 0.10  & 0.10 & 2.0 & 26.6578 & 0.0715111 & 20.0284 & 0.0537274   & 1.13104 & -0.413689 & 2.63043\\
                      & 0.10  & 0.10 & $\infty$ & 24.0469 & 0.0530149 & 18.0668 & 0.0398309  & 1.09672 & -0.425657 & 2.37281\\
\hline
\multirow{4}{*}{0.1}   & 0.25  & 0.25 & 1.2 & 36.58 & 0.222988 & 27.4831 & 0.113963  & 1.28608 & -0.276002 & 3.6095  \\
                      & 0.25  & 0.25 & 1.4 & 30.0954 & 0.132432 & 22.6111 & 0.0994979  & 1.22571 & -0.367598 & 2.96964  \\
                      & 0.25  & 0.25 & 2.0 & 25.5194 & 0.0841841 & 19.1731 & 0.0632488  & 1.17203 & -0.414005 & 2.5181 \\
                      & 0.25  & 0.25 & $\infty$ & 22.9847 &0.062994 & 17.2688 & 0.0473283 & 1.1375 & -0.427547 & 2.268\\
\hline
\multirow{4}{*}{0.3}  & 0.10  & 0.10 & 1.2 & 31.565 & 0.494733 & 23.7153 & 0.3717  & 1.61897 & -0.444896 & 3.11465 \\
                      & 0.10  & 0.10 & 1.4 & 25.6834 & 0.318908 & 19.2963 & 0.239601 & 1.55668 & -0.548617 & 2.53428 \\
                      & 0.10  & 0.10 & 2.0 & 21.5156 & 0.218539 & 16.165 & 0.164192  & 1.49933 & -0.598094 & 2.12304\\
                      & 0.10  & 0.10 & $\infty$ & 19.1999 & 0.172071 & 14.4252 & 0.129279  & 1.4616 & -0.6079 & 1.89454 \\
\hline
\multirow{4}{*}{0.3}   & 0.25  & 0.25 & 1.2 & 29.4773 & 0.638265 & 22.1467 & 0.479538  & 1.82734 & -0.720304 & 2.90865 \\
                      & 0.25  & 0.25 & 1.4 & 23.8921 & 0.428647 & 17.9505 & 0.322049  & 1.77993 &-0.873316 & 2.35753\\
                      & 0.25  & 0.25 & 2.0 & 19.9254 & 0.307738 & 14.9703 & 0.231208 & 1.74127 & -0.978784 & 1.96612 \\
                      & 0.25  & 0.25 & $\infty$ & 17.7162 & 0.251906 & 13.3104 & 0.189261  & 1.72008 & -1.03261 & 1.74813  \\
\hline\hline
	\end{tabular}

}	
\end{table*}

Here, $\theta_\infty$ is the angular position acquired by the set of images in the limit $n \to \infty $ or in other words angular radius of photon sphere. s is the angular seperation between outermost image ($n=1$) and innermost packed images ($n=2,3,4...\infty$).

The EHT using VLBI technique, released the first image of supermassive black hole M87* in the center of nearby elliptical galaxy. The image (shadow) is an asymmetric emission ring with diameter of $42\pm3 \mu$as and indicated mass of the black hole $(6.5 \pm 0.7) \times 10^9 M_{\odot}$. This offers a new tool to test black hole in the strong field regime.  Motivated by this we  consider the  supermassive black holes Sgr A* in our galactic center and M87* in nearby galaxy Messier 87 as lens for numerical estimation of observables.  The masses and distances from earth according to the latest observational data for Sgr A* \cite{Do:2019vob} are $M = 4.3 \times 10^6 M_{\odot}$, $D_{OL} = 8.35$ Kpc and for  M87* \cite{Akiyama:2019vo} are  $M = (6.5\pm 0.7) \times 10^9 M_{\odot}$, $D_{OL} = (16.8\pm0.8)$ Mpc respectively. We have tabulated the strong lensing coefficients $\bar{a}$, $\bar{b}$, $u_m$ and the observables angular position of innermost image $\theta_{\infty}$ and separation between the first image and the inner packed ones  $s$  in Table~\ref{table1}. For comparisons, we have estimated these values for the Schwarzschild black hole and Kerr-Newman black hole and found that the angular position and angular separation for  EBI black hole are larger. In Table~\ref{table6}, we have calculated the deviation of the coefficients and observables of the EBI black hole from the Kerr black hole for fixed value of spin parameter ($a=0.3$) and found that in case of EBI black hole the images are formed closer to the optical axis but are more packed than the Kerr black hole. The deviation of angular positions can reach a maximum value of $2.5\mu$as for Sgr A* and $1.8\mu$as for M87*, which with the current observational facilities is impossible to resolve. From the plots in Figs.~(\ref{plot5}) and ~(\ref{plot6}) it can be immediately followed that angular separation $s$ increases but angular position ($\theta_{\infty}$)  decreases with spin parameter $a$. It is worth noting that both angular separation and angular position increase in the presence of plasma than in vacuum. Also, for fixed values of $a$ and $\tilde{E}$, both angular position and angular separation decreases with charge $q$.

\begin{table*}[t]
\caption{Deviation of the lensing observables of EBI black holes from Kerr black hole by taking Sgr A* and M87* as lens, where $\delta X = X_{\text{Kerr}}-X_{\text{EBI}} $\label{table6}}
\resizebox{1\textwidth}{!}{

	\begin{tabular}{p{0.8cm} p{0.8cm} p{0.8cm} p{1cm} p{1.5cm} p{2.0cm} p{1.5cm} p{2.0cm} p{1.5cm} p{1.5cm} p{1.2cm}}
		
\hline\hline
\multicolumn{4}{c}{}&
 \multicolumn{2}{c}{Sgr A*} & \multicolumn{2}{c}{M87*}  & \multicolumn{2}{c}{Lensing Coefficients}\\
{$a$ } & {$q$} & {$\beta$} & {$\tilde{E}$} & {$\delta\theta_\infty $ ($\mu$as)} & {$\delta s$ ($\mu$as) } &    {$\delta \theta_\infty $ ($\mu$as)} & {$\delta s$ ($\mu$as) } &   {$\delta \bar{a}$}&{$\delta \bar{b}$} & {$\delta u_m/R_s$}\\
\hline

\multirow{4}{*}{0.3}   & 0.25  & 0.25 & 1.2 &  2.4438 & -0.163407 & 1.83607 & -0.12277  & -0.231673 & 0.288928 & 0.24114  \\
                      & 0.25  & 0.25 & 1.4  &  2.09623 & -0.124046 & 1.57493 & -0.0931977  &  -0.247045 & 0.340692 & 0.206844\\
                      & 0.25  & 0.25 & 2.0 & 1.85985 & -0.100026 & 1.39733 & -0.075151 & -0.266229 & 0.398848 & 0.183519  \\
                      & 0.25  & 0.25 & $\infty$ & 1.73419 & -0.088959 & 1.30292 & -0.0668363 &  -0.283155 & 0.444148 & 0.17112  \\
\hline
	\end{tabular}
}	
\end{table*}

\section{Weak-field lensing}\label{lensing}
In this section we review the effects of gravitational lensing in the background of EBI black hole surrounded by a plasma
considering a weak-field approximation defined as follows,

\begin{figure*}[t]
 \begin{center}
   \includegraphics[scale=0.3]{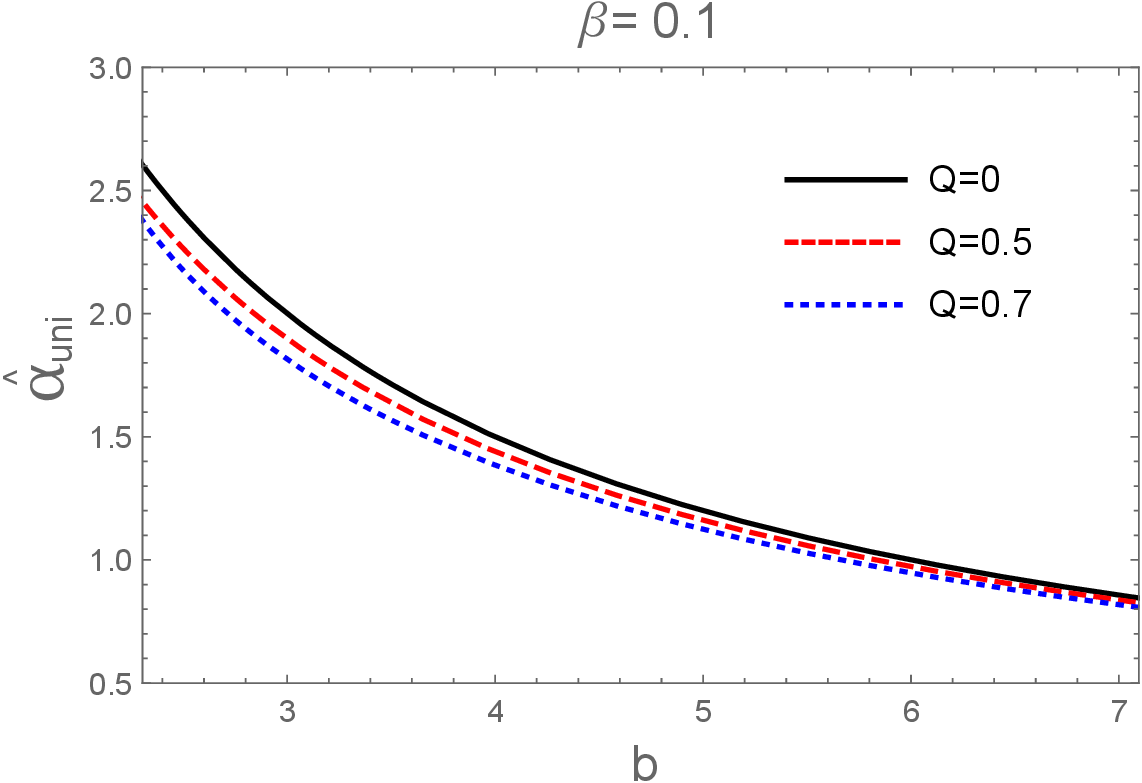}
   \includegraphics[scale=0.3]{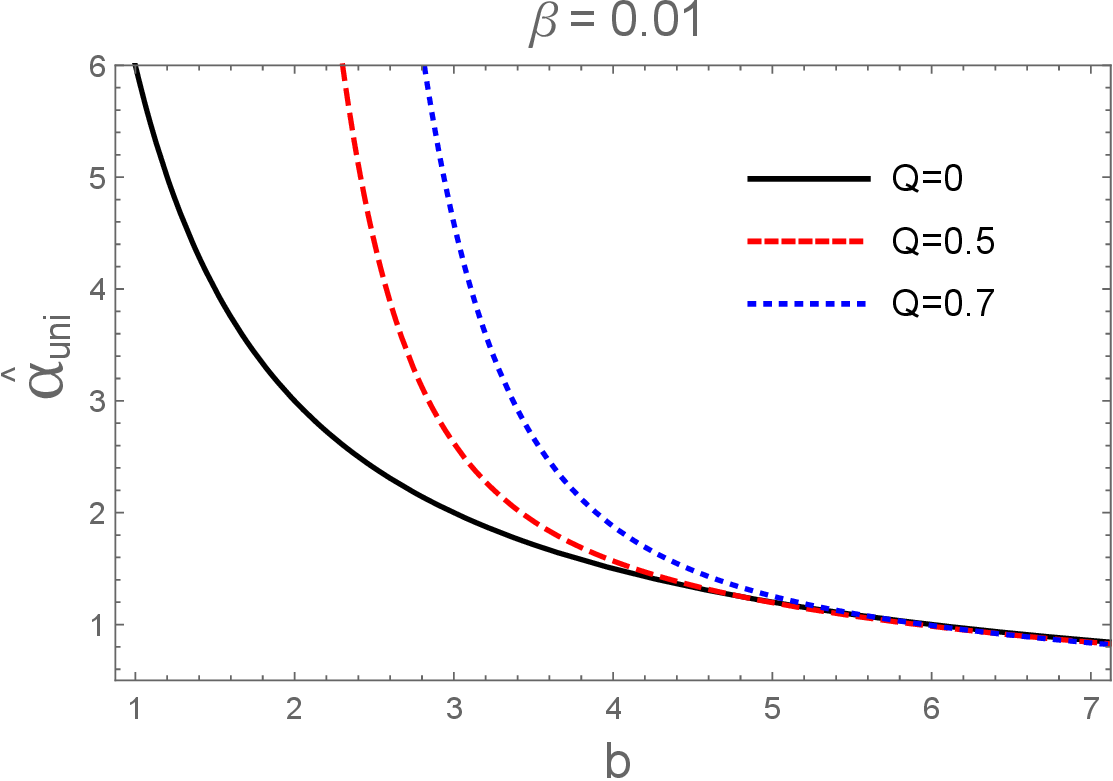}
   \includegraphics[scale=0.3]{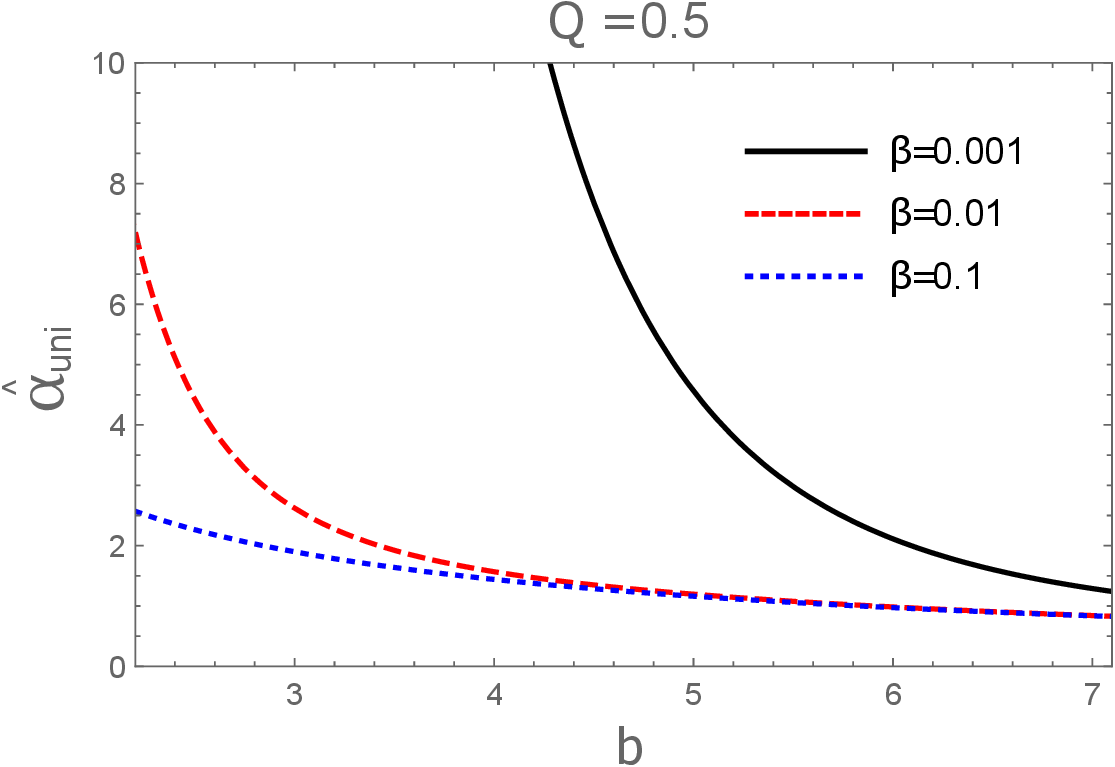}

    \includegraphics[scale=0.3]{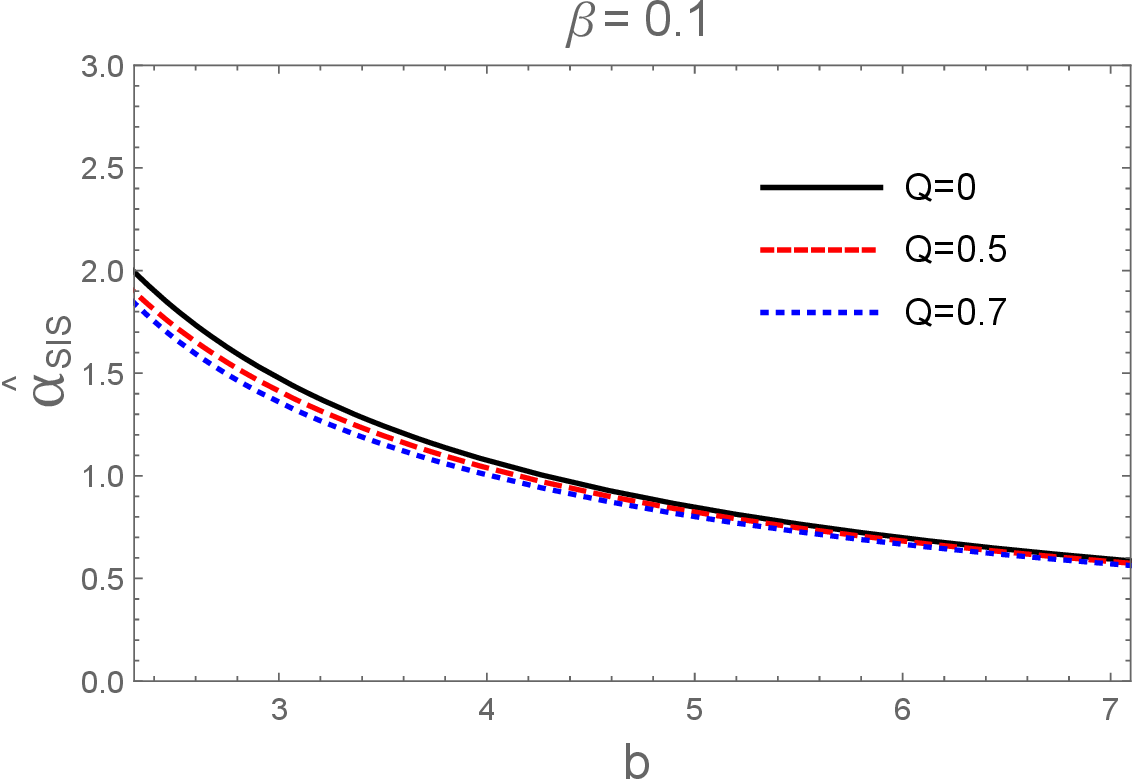}
   \includegraphics[scale=0.3]{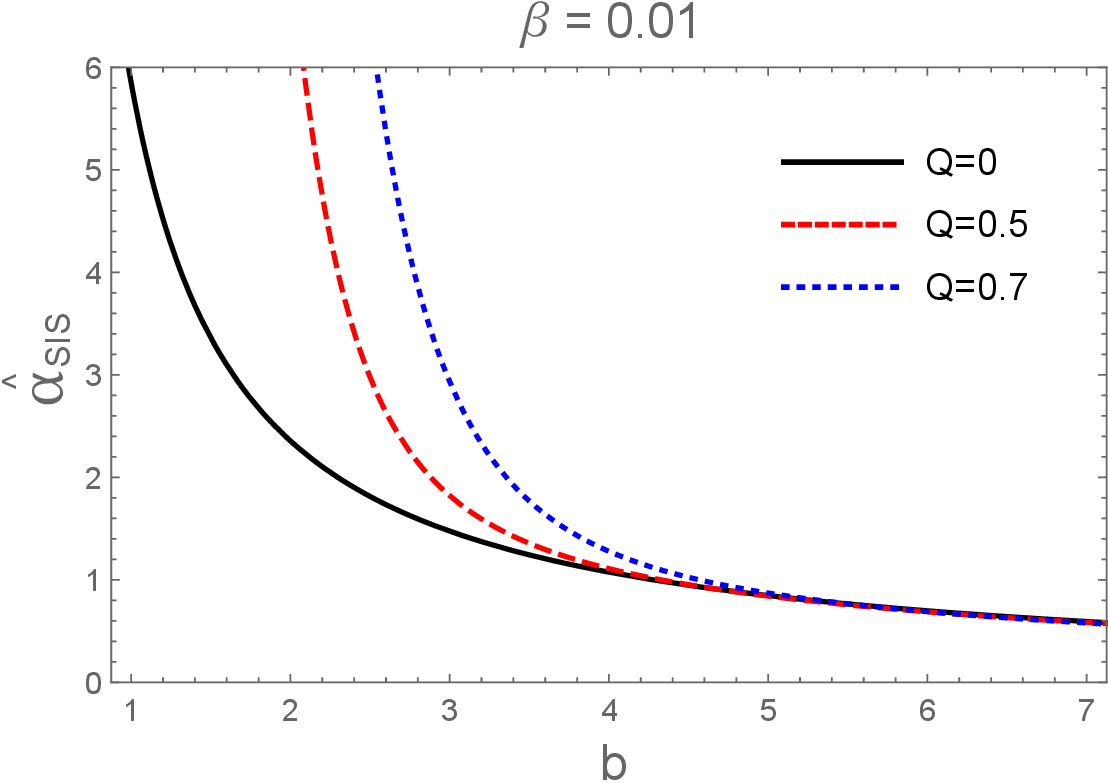}
   \includegraphics[scale=0.3]{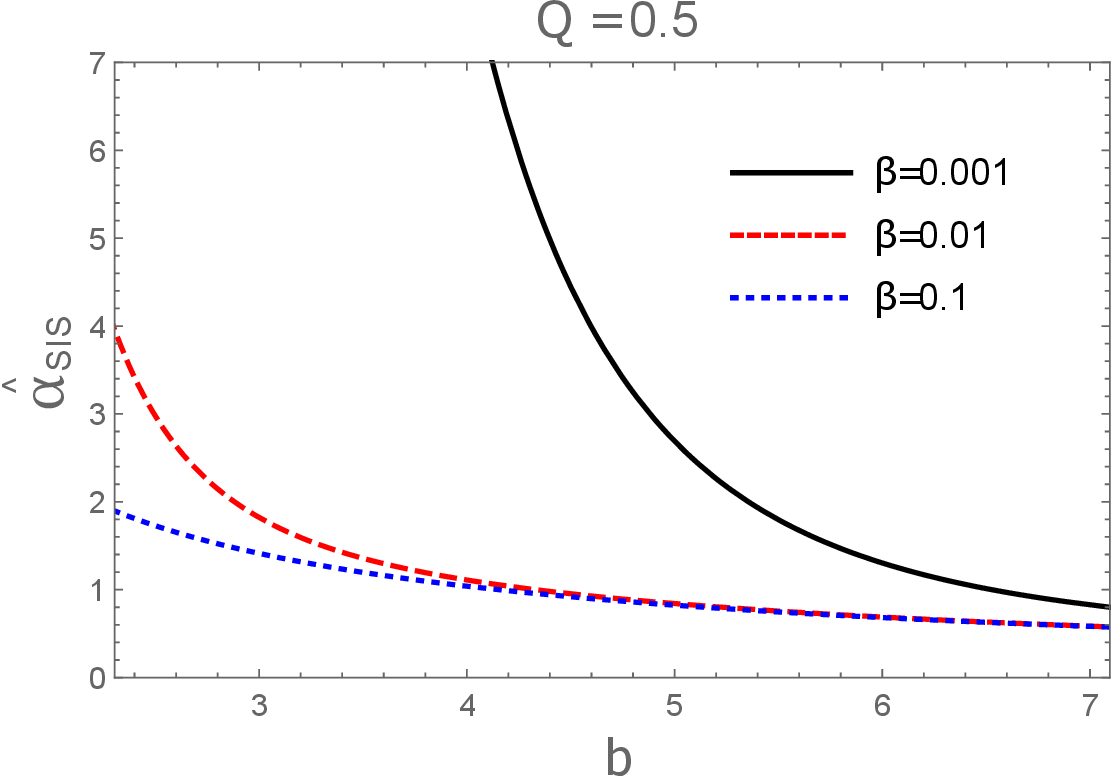}
  \end{center}
\caption{Deflection angle $\hat{\alpha}_{b}$ as a function of the impact
parameter $b$ for different charge $Q$ and Born-Infeld $\beta$ values for fixed $\frac{\omega^2_{e}}{\omega^2}=0.5$(upper panel).
Deflection angle $\hat{\alpha}_{SIS}$ as a function of the impact
parameter $b$ for different charge $Q$ and Born-Infeld $\beta$ values for fixed $\frac{\omega^2_{c}}{\omega^2}=0.5$ (lower panel).}\label{Anglea}
\end{figure*}

\begin{align}
g_{\alpha\beta}=\eta_{\alpha\beta}+h_{\alpha\beta},
\end{align}
\\
where $\eta_{\alpha\beta}$ and $h_{\alpha\beta}$ specify the Minkowski metric and perturbation metric, respectively.
\begin{align}
\eta_{\alpha\beta}&={\rm diag} (-1, 1, 1, 1)\ ,\nonumber\\
 h_{\alpha\beta} &\ll 1, \quad  h_{\alpha\beta} \rightarrow 0 \quad {\rm under } \quad x^{\alpha}\rightarrow \infty\ , \nonumber \\
 g^{\alpha\beta}&=\eta^{\alpha\beta}-h^{\alpha\beta},\ \ \ h^{\alpha\beta}=h_{\alpha\beta}\ .
\end{align}
\\
We follow the notion in \cite{Abu:2017aa} by taking into account the weak-field approximation and weak plasma strength for the photon propagation
along $z$ axis to get the angle of deflection,

\begin{align}
\hat{\alpha}_i &= \frac{1}{2} \int_{-\infty}^\infty \bigg(h_{33_,i} +
\frac{ \omega^2}{\omega^2-\omega^2_{e}}h_{00_,i} -\frac{K_e }{\omega^2-\omega^2_{e}}N{_,i}\bigg) dz. \label{alphak}
\end{align}
\\
Note that, the $\hat{\alpha}_i<0$ and  $\hat{\alpha}_i>0$ indicate the deflection
towards and away from the central object, respectively. In the last expression the values
of $\omega_e$ and $n$ are set to limitations at infinity and $\omega(\infty)=\omega$.
At large $r$, the black hole metric could be approximated to \cite{Bin:2010a}.

\begin{align}
ds^2 = ds^2_{0}+\left(\frac{2M}{r}-\frac{Q^2(r)}{r^2}\right)dt^2+\left(\frac{2M}{r}-\frac{Q^2(r)}{r^2}\right) dr^2, \label{metrbb}
\end{align}
\\
where
$ds^2_{0}=-dt^2+dr^2+r^2(d\theta^2+\sin^2\theta d\phi^2)$.
In the Cartesian coordinates the components $h_{\alpha\beta}$ can be written as

\begin{align}
 h_{00}=&\left(\frac{R_g}{r}-\frac{Q^2(r)}{r^2}\right), \nonumber \\  h_{ik}=&\left(\frac{R_g}{r}-\frac{Q^2(r)}{r^2}\right)n_{i}n_{k}\ ,\nonumber \\
 h_{33}=&\left(\frac{R_g}{r}-\frac{Q^2(r)}{r^2}\right)\cos^2x ,\label{h}
\end{align}
where $R_g=2M$, $\cos x=z/\sqrt{b^2+z^2}$ and $r=\sqrt{b^2+z^2}$. We apply the approximation $\beta\rightarrow\infty$ to the series expansion of $Q^2(r)$ (\ref{Q(r)a}),
\begin{align}
Q^2(r)= Q^2 - \frac{2 Q^4}{15 r^4 \beta^2}.
\end{align}
\\
Using the above mentioned expressions in the formula (\ref{alphak}) one can compute the light deflection angle
for a black hole in plasma in terms of the impact parameter $b$, given as below

\begin{align}
\hat{\alpha}_{b}&=\bigintssss_{-\infty}^{\infty}\frac{b}{2r}
\Bigg(\partial_r \bigg(\bigg(\frac{R_g}{r}-\frac{Q^2(r)}{r^2}\bigg)\cos^2x\bigg)\nonumber \\&+
\partial_r\bigg(\frac{R_g}{r}-\frac{Q^2(r)}{r^2}\bigg)\frac{\omega^2}{\omega^2-\omega^2_{e}}
-\frac{K_e}{\omega^2-\omega^2_{e}}\partial_r N\Bigg)dz,
\label{alfa}
\end{align}
\subsection{Uniform Plasma}
Now, we discuss the light deflection in the presence of a uniform plasma,
thereby considering $\omega_e$ as a constant quantity and taking in the approximation $1-n<<\frac{\omega_{e}}{\omega}$.
 As a result $\partial_r N$ vanishes and the angle of deflection takes the form
\begin{align}
\hat{\alpha}_\mathrm{uni}&=-\bigg(\frac{\omega^2}{\omega^2-\omega^2_{e}}\bigg)\bigg(\frac{\pi Q^2}{2b^2}-\frac{\pi Q^4}{8b^6\beta^2}-\frac{R_g}{b}\bigg)
\nonumber \\&-\bigg(\frac{\pi Q^2}{4b^2}-\frac{\pi Q^4}{48b^6\beta^2}-\frac{R_g}{b}\bigg).
\end{align}
The upper panel of Fig.~(\ref{Anglea}) illustrates the angle of deflection as a function of the impact parameter.
On varying the charge for a fixed $\beta$, the angle of deflection executes two opposite
behaviours depending on the value of $\beta$. The deviation of photons is larger as the value of $Q$ decreases
when allotted with a comparatively larger $\beta$, hence, $\hat{\alpha}_\mathrm{uni}$ is maximum for $Q=0$, which basically refers to the
Schwarzschild gravity. It is noticed that by increasing $\beta$ for a fixed charge $Q$ the results are consistent, and the angle of deflection reduces in any case.

\subsection{Singular isothermal sphere}
We shall now examine the photon deflection by the black hole surrounded by a singular isothermal sphere,
which was modeled in \cite{Chnd:1939a,Jbin:1987a} and has up till now played
an important part to explore the len's property of the galaxies and clusters.
The density distribution has the form

\begin{align}
\rho(r)=\frac{\sigma^2_{v}}{2\pi r^2},\label{rho}
\end{align}
\\
where $\sigma^2_{v}$ is a one-dimensional velocity dispersion. The
concentration of the plasma has the form

\begin{align}
N(r)=\frac{\rho(r)}{\kappa m_p},\label{conelec}
\end{align}
\\
where $m_p$ is the proton mass and $\kappa$ is a non-dimensional
coefficient which is related to the dark matter contribution. Utilizing (\ref{omegae},\ref{rho},\ref{conelec}) the plasma frequency takes the form

\begin{align}
\omega^2_{e}=K_eN(r)=\frac{K_e\sigma^2_{v}}{2\pi \kappa m_p r^2}.
\end{align}
\\
The angle of deflection using the latter particulars can be conveniently computed as below \cite{Bin:2010a},

\begin{align}
\hat{\alpha}_\mathrm{SIS}&=\frac{\omega^2_{c}R^2_{g}}{\omega^2b^2}\bigg(\frac{1}{2}-\frac{3Q^2}{8b^2}
+\frac{7Q^4}{64b^6\beta^2}+\frac{2R{_g}}{3b\pi}\bigg) -\nonumber \\&\bigg(\frac{3 \pi Q^2}{4b^2}-\frac{7\pi Q^4}{48b^6\beta^2}-\frac{2R_g}{b}\bigg),
\end{align}
we get $\omega^2_{c}$ which is another plasma constant and has the following analytic expression.
\begin{align}
\omega^2_{c}=\frac{\sigma^2_{v} K_e}{2\kappa m_p R^2_{g}}.
\end{align}
\\
The lower panel of Fig.~(\ref{Anglea}) shows the angle of deflection $\hat{\alpha}_\mathrm{SIS}$ as a function of the impact parameter
considering a black hole veiled by a singular isothermal sphere. The results obtained for distinct $Q$ and $\beta$ values,
altogether share a common thread with the uniform plasma case, but, however the angle of deflection is relatively smaller as by the
black hole surrounded with a uniform plasma, i.e, $\hat{\alpha}_\mathrm{uni}>\hat{\alpha}_\mathrm{SIS}$, see details in \cite{Babar:2021a}.

\section{Conclusion} \label{con}
In this paper we constructed an insightful discussion regarding the rotating black hole's structure,
centre-of-mass energy and gravitational lensing in Einstein-Born-Infeld gravity.
It is searched out that the radii of the event horizons, the static limit surfaces and ISCOs experience
a uniform decrease as the strength of black hole's charge intensifies but the Cauchy horizon
on the contrary, exhibits an increase. Furthermore, our investigation
ensures the BSW effect in the EBI space-time, henceforth the centre-of-mass energy diverges near the
horizon just like the other charged rotating black holes, in our case we specifically considered the Kerr-Newman black hole.

The gravitational lensed photons considering a uniform plasma and a singular isothermal sphere are also part of the context.
Unfortunately due to extinction of radiation in the vicinity of galactic center, the observation of relativistic images is an uphill task. In addition, the radiation from the accreting materials badly influences the observation of these images. Compared to the weak lensing, these obstacles are even bigger for the strong lensing. For relativistic images to be more prominent, the lens components (the source, the observer and the lens) should be highly aligned. A suitable source for the strong lensing could be a supernova but the probability of it to be aligned with the lens and observer is extremely small. Despite this, there is no doubt that observation of relativistic images would be one of the most important discovery in the field of astronomy and would have immense implication for general relativity and relativistic physics, one of which is that it could provide a test for general relativity in the strong field regime.

In this paper it is interpreted that the photons deviate at a larger angle when a uniform plasma walls in the black hole. The light deflection coefficients $\bar{a}$ and $\bar{b}$, in the strong field limits, and their variance with the rotational parameter $a$ for different plasma frequency as well as in vacuum are calculated. The effect of plasma result in the increase of the photon sphere radius, the deflection angle and  the strong deflection coefficients $\bar{a}$ and $\bar{b}$ as compared to the Kerr-Newman black hole. The lensing observables angular positions and the angular separation between the relativistic images show similar behavior. It is also shown in the paper that with increasing spin the impact of plasma on strong gravitational lensing becomes smaller as the spin parameter increases in the prograde orbit ($a>0$) especially for the case of an extreme black hole, the strong gravitational effects in the homogenous plasma are same as the case in vacuum for $a>0$. Also, in our analysis, the EBI gravity has sustained the regularity of a charged black hole effectively in all aspects. In reality, the plasma can be significantly non-uniform in the close vicinity of compact objects  like black holes. Such cases, though complicated, will be considered in our later research.

\section*{Acknowledgements} F.A. acknowledges the support of INHA
University in Tashkent. S.G.G. and S.U.I would like to thank SERB-DST for the ASEAN project IMRC/ AISTDF/ CRD/ 2018/ 000042. S.G.G would also like to thank IUCAA, Pune for the hospitality while this work was being done.

\section{Appendix A: Calculting deflection angle in Strong Deflection Limit} \label{sect7}
In this section we describe the details of calculation for the deflection angle in strong deflection limit by EBI black hole using the approach discussed in \cite{Bozza:2001aa}. The metric for EBI black hole when both observer and source are in equatorial plane is given by
\begin{equation}\label{metric1}
ds^2 = -A(x)dt^2 + B(x)dx^2 + C(x)d\phi^2- 2 D(x)dtd\phi,
\end{equation}
\\
where

\begin{eqnarray}
A(x) &=& \frac{\Delta-a^2}{x^2},\\
B(x) &=& \frac{x^2}{\Delta}, \\
C(x) &=& x^2+a^2\left(2-\frac{\Delta-a^2 }{x^2}\right),\\
D(x) &=& a\left(1-\frac{\Delta-a^2 }{x^2}\right),
\end{eqnarray}
and $\Delta=x^2-2Mx+a^2+q^2(x)$,

\begin{align} \label{qx}
 q^2(x)&\approx\frac{2\beta'^2x^4}{3}\bigg(1-\sqrt{1+\xi^2(x)}\bigg),
 \nonumber \\&+\frac{4 q^2}{3}\bigg(1 -\frac{1}{10}\xi^2(x)+\frac{1}{24}\xi^4(x)\bigg),
\end{align}
\\
Here $x$ is in the unit of Schwarzschild radius. Using  Eq.(\ref{pdot}),  Eq.(\ref{rdot}) we have

\begin{eqnarray}
\frac{d\phi}{dx} &=& \frac{\sqrt{B(x)A(x_0)}(\tilde{E}D(x)+\tilde{L}A(x))}{\sqrt{D(x)^2+A(x)C(x}\sqrt{P(x,x_0)}},
\end{eqnarray}
\begin{eqnarray}
P(x,x_0)&&= G(x)A(x_0)-G(x_0)A(x)+2\tilde{E}\tilde{L}(A(x)D(x_0)\nonumber\\&& -A(x_0)D(x)),
\end{eqnarray}
\\
where $x_0$ is the distance of closest approach to the black hole.
The deflection angle will be

\begin{eqnarray}\label{def5}
\alpha_D(x_0) = 2 \int_{x_0}^{\infty}\frac{d\phi}{dx}dx
-\pi \equiv  I_{T}(x_0) - \pi.
\end{eqnarray}
\\
To solve the integral (\ref{def5}), we define a variable
\begin{equation}
z = 1-\frac{x_0}{x},
\end{equation}
\\
such that the integral (\ref{def5}) reduces to
\begin{eqnarray}\label{def6}
I_T(x_0) &=& \int_{0}^{1}R(z,x_0) f(z,x_0) dz,
\end{eqnarray}
\\
with the functions
\begin{eqnarray}
R(z,x_0) &=& \frac{\sqrt{B(x)A(x_0)}(\tilde{E}D(x)+\tilde{L}A(x))}{\sqrt{D(x)^2+A(x)C(x}},\label{RF}\\
f(z,x_0) &=& \frac{1}{\sqrt{A(x_0)-\frac{A(x)}{C(x)}C(x_0)}},\label{RF1}
\end{eqnarray}
\\
where $x = \frac{x_0}{1+z}$.
Taking Taylor expansion of the expression under the square root in Eq.~(\ref{RF1}) we have
\begin{eqnarray}
f_0(z,x_0) &=& \frac{1}{\sqrt{\zeta(x_0) z + \eta(x_0) z^2}},
\end{eqnarray}
\\
where
\begin{eqnarray}\label{zeta}
\zeta(x_0) &=& \frac{x_0}{C(x_0)}\left[C^\prime(x_0) A(x_0)-A^\prime(x_0) C(x_0)\right],
\end{eqnarray}

\begin{eqnarray}\label{eta}
\eta(x_0) &=& \frac{1}{2  C(x_0)^2}\Big[2x_0 C(x_0)(A(x_0)C^\prime(x_0)-A^\prime(x_0)C(x_0))  \nonumber\\
&+&  2x_0^2 (C(x_0)A^{\prime}(x_0)C^{\prime}(x_0)-A(x_0)C^{\prime}(x_0)^2) \nonumber\\
&& -x_0^2C(x_0)(C(x_0)A^{\prime\prime}(x_0) -A(x_0)C^{\prime\prime}(x_0) )\Big].
\end{eqnarray}
\\
The radius of photon sphere $x_m$ is given by solving $\zeta(x_0)=0$.
In the integral (\ref{def6}) $R(z,x_0)$ is regular everywhere but as $x_0=x_m$, $\zeta(x_0)=0$ and $f(z,x_0)\approx 1/z$, which diverges as $z \to 0$. Thus we separate the integral (\ref{def6}) into two parts as
\begin{eqnarray}
I(x_0) = I_D(x_0) + I_R(x_0),
\end{eqnarray}
\\
where
\begin{equation}\label{divergent}
I_D(x_0)=\int_0 ^1 R(0,x_m)f_0(z,x_0)dz,
\end{equation}
\\
is the divergent part and regular part is
\begin{eqnarray}\label{regular}
I_R(x_0)&=& \int_0 ^1 \Big(R(z,x_0)f(z,x_0)\nonumber \\ &-&R(0,x_m)f_0(z,x_0)\Big)dz.
\end{eqnarray}
\\
For EBI black hole Eq.~(\ref{divergent}) can be solved analytically to
\begin{eqnarray}
I_D(x_0) &=& -\frac{R(0,x_m) }{\sqrt{\eta(x_0)}} \log\Big(\frac{\sqrt{\zeta(x_0)+\eta(x_0)}+\sqrt{\eta(x_0)}}{\sqrt{\eta(x_0)}}\Big) \nonumber\\&& + O(x_0-x_m),
\end{eqnarray}
\\
And regular term can be solved numerically as
\begin{equation}
I_R(x_0)=I_R(x_m) + O(x_0-x_m).
\end{equation}
\\
By expanding the Eq.~(\ref{zeta}) and Eq.~(\ref{eta}) about $x_m$, the deflection angle in SDL becomes
\begin{equation}\label{deff2}
\alpha_D(x_0)=-a \log\Big(\frac{x_0}{x_m}-1\Big)+ b + O(x_0-x_m),
\end{equation}
where
\begin{eqnarray}\label{ab2}
a &=& \frac{R(0,x_m) }{\sqrt{\eta(x_m)}},\\
b &=& -\pi + I_R(x_m).
\end{eqnarray}
\\
The impact parameter can be expressed as
\begin{equation}\label{impact}
u = \frac{\tilde{L}(x0)}{\sqrt{\tilde{E}^2-1}}.
\end{equation}
\\
We can write deflection angle in terms of impact parameter by expanding Eq.~(\ref{impact}) around $x_m$ as,
\begin{eqnarray}\label{defangle}
\alpha_D(u) &=& -\bar{a} \log\Big(\frac{u}{u_m}-1\Big)+\bar{b} + O(u-u_m),
\end{eqnarray}
\\
where the strong deflection coefficients are
\begin{equation}
\bar{a}=\frac{a}{2},
\end{equation}
\begin{equation}
\bar{b}= b +\bar{a} \log \left(  \frac{\eta(x_m) G(x_m)}{u_m\sqrt{\tilde{E^2}-1}A(x_m)(\tilde{E}D(x_m) + \tilde{L} A(x_m)) }\right),
\end{equation}
\begin{equation}
u_m = \frac{-\tilde{E}D(x_m) + \sqrt{A(x_m) G(x_m)+\tilde{E}^2 D(x_m)^2}}{\sqrt{\tilde{E}^2-1} A(x_m)}.
\end{equation}

\end{document}